\documentclass[10pt,twocolumn,aps,pra,showpacs,superscriptaddress,floatfix,longbibliography,nofootinbib]{revtex4-2}

\usepackage{graphicx}
\usepackage{subfigure}
\usepackage[T1]{fontenc}
\usepackage{graphicx}
\usepackage[utf8]{inputenc}
\usepackage{amsmath, amsthm, amssymb,amsfonts}
\usepackage{color}
\usepackage{psfrag}
\usepackage{epsfig}
\usepackage{bbm}
\usepackage{bm}
\usepackage[colorlinks=true,citecolor=blue,urlcolor=magenta]{hyperref}
\usepackage[normalem]{ulem}
\usepackage{epstopdf}
\usepackage{graphicx}
\usepackage{stackengine,xcolor}
\usepackage{comment}

\usepackage{tikz}
\usetikzlibrary{decorations.pathreplacing}

\definecolor{nred}{rgb}{0.9,0.1,0.1}
\definecolor{nblack}{rgb}{0,0,0}
\definecolor{nblue}{rgb}{0.2,0.2,0.8}
\definecolor{ngreen}{rgb}{0.2,0.6,0.2}

\usepackage{etoolbox}
\usepackage{bbold}

\newcommand{\ket}[1]{| #1 \rangle}
\newcommand{\bra}[1]{\langle #1 |}
\newcommand{\proj}[1]{\ket{#1}\!\bra{#1}}

\newcommand{\ketbrac}[2]{|#1\rangle\langle #2|}

\newcommand{\rb}[1]{\left( #1 \right)}

\newcommand{\ew}[1]{\langle #1 \rangle}
\newcommand{\beq}{\begin{eqnarray}}
\newcommand{\eeq}{\end{eqnarray}}

\newcommand{\op}[2]{| #1 \rangle \langle #2 |}

\newcommand{\eq}[1]{Eq.~(\ref{#1})}
\newcommand{\fig}[1]{Fig.~\ref{#1}}

\newcommand{\id}{\openone}

\newcommand{\M}{\mathcal{M}}
\newcommand{\Q}{\mathcal{Q}}

\newcommand{\MAA}{\{A_{a|x}\}_{x,a}}
\newcommand{\MAB}{\{B_{b|y}\}_{y,b}}

\newcommand{\SAB}{\{\rho_{a|x}\}}

\newcommand{\tA}{\text{A}}
\newcommand{\tB}{\text{B}}
\newcommand{\tC}{\text{C}}

\newcommand{\ttC}{\text{\tiny C}}
\newcommand{\tAB}{\text{AB}}
\newcommand{\ttAB}{\text{\tiny AB}}

\newcommand{\ttAC}{\text{\tiny AC}}
\newcommand{\ttBC}{\text{\tiny BC}}
\newcommand{\rbc}{{\rho_{\ttBC}}}
\newcommand{\raxbc}{{\rho_{a|x}^{\ttBC}}}
\newcommand{\chiDIbc}{\chi_{\mbox{\tiny DI}}^{(\ell)}[\raxbc]}
\newcommand{\chiDIbcfirst}{\chi_{\mbox{\tiny DI}}^{(1)}[\raxbc]}
\newcommand{\pobs}[1]{P_{\mbox{\tiny obs}}(#1)}
\newcommand{\pobsAB}[1]{P_{\mbox{\tiny obs}}^{\ttAB}(#1)}
\newcommand{\pobsAC}[1]{P_{\mbox{\tiny obs}}^{\ttAC}(#1)}
\newcommand{\SR}{\mathrm{SR}}

\newcommand{\tBC}{\text{BC}}
\newcommand{\Bobs}{\mathcal{B}_{\mbox{\tiny obs}}}
\newcommand{\IR}{\mathrm{IR}}
\newcommand{\SRDItwoone}{\SR_{\mbox{\tiny DI},\ell}^{\,\mbox{\tiny L},\tAB\to\tC}}
\newcommand{\SRDINStwoone}{\SR_{\mbox{\tiny DI},\ell}^{\,\mbox{\tiny NS},\tAB\to\tC}}
\newcommand{\SRDIonetwo}{\SR_{\mbox{\tiny DI},\ell}^{\,\tA\to\tBC}}

\newcommand{\rax}{{\rho_{a|x}}}

\newcommand{\rc}{{\rho_\ttC}}
\newcommand{\rabc}{{\rho_{\mbox{\tiny ABC}}}}

\newcommand{\rabxy}{{\rho_{ab|xy}^{\mbox{\tiny C}}}}
\newcommand{\vecP}{\mathbf{P}}
\newcommand{\Pobs}{\mathbf{P}_\text{obs}}
\newcommand{\Pobsabcxyz}{{P_{\mbox{\tiny obs}}(abc|xyz)}}

\newcommand{\chil}{\chi^{(\ell)}}
\newcommand{\chilbb}{\chi_{\mbox{\tiny DI}}^{(\ell)}}
\newcommand{\chiDI}{\chi_{\mbox{\tiny DI}}^{(\ell)}[\rho_{a|x}]}
\newcommand{\chiDIfirst}{\chi_{\mbox{\tiny DI}}^{(1)}[\rho_{a|x}]}
\newcommand{\chiDIabc}{\chi_{\mbox{\tiny DI}}^{(\ell)}[\rho_{ab|xy}^{\mbox{\tiny C}}]}
\newcommand{\chiDIabcfirst}{\chi_{\mbox{\tiny DI}}^{(1)}[\rho_{ab|xy}^{\mbox{\tiny C}}]}

\DeclareMathOperator{\tr}{tr}

\theoremstyle{definition}

\newtheorem{definition}{Definition}
\newtheorem{lemma}{Lemma}

\newtheorem{proposition}{Proposition}
\newtheorem{corollary}{Corollary}

\newcommand{\SRL}{\SR^{\mbox{\tiny L}}}
\newcommand{\SRQ}{\SR^{\mbox{\tiny Q}}}
\newcommand{\SRNS}{\SR^{\mbox{\tiny NS}}}
\newcommand{\PNS}{P_{\mbox{\tiny NS}}}
\newcommand{\IRNS}{\IR^{\mbox{\tiny NS}}}
\newcommand{\MAAB}{\{A_{a|x}\otimes B_{b|y}\}_{x,y,a,b}}

\begin{document}

\title{Device-independent quantification of steerability in tripartite scenario}

\author{Xin-Hong Wang}
\thanks{These authors contributed equally to this work.}
\affiliation{Department of Physics, National Chung Hsing University, Taichung 40227, Taiwan}

\author{Miao-Tzu Lin}
\thanks{These authors contributed equally to this work.}
\affiliation{Department of Physics, National Chung Hsing University, Taichung 40227, Taiwan}

\author{Shin-Liang Chen}
\email{shin.liang.chen@email.nchu.edu.tw}
\affiliation{Department of Physics, National Chung Hsing University, Taichung 40227, Taiwan}
\affiliation{Physics Division, National Center for Theoretical Sciences, Taipei 106319, Taiwan}
\affiliation{Center for Quantum Frontiers of Research \& Technology (QFort), National Cheng Kung University, Tainan 701, Taiwan}

\date{ \today}

\begin{abstract}
Quantum steering captures the ability of one party, through local measurements on a shared entangled state, to affect the conditional states held by distant parties in a way that admits no local explanation. Its device-independent (DI) quantification---requiring no characterization of any state or measurement---has been developed in the bipartite setting, notably through the framework of assemblage moment matrices (AMMs) [Phys. Rev. Lett. \textbf{116}, 240401 (2016)], but has remained largely unexplored beyond it. Here, we extend the AMM framework to tripartite steering, covering both the scenario in which two parties jointly steer a third one (2-steer-1) and that in which a single party steers the remaining two (1-steer-2). For each scheme, we construct a semidefinite program that, from the observed correlations---or merely from the violation of a Bell inequality---yields lower bounds on the steering robustness of the underlying assemblage. As a by-product, our bounds also certify, in a DI manner, the incompatibility of the measurements performed by the steering parties.
\end{abstract}

\maketitle

\section{Introduction}
\label{Sec_intro}
Local measurements on the parts of an entangled quantum state can generate correlations that exclude any explanation in terms of local causes. This phenomenon, known as Bell nonlocality~\cite{Bell64,NLreview}, has by now been confirmed in loophole-free experiments~\cite{Hensen15,Giustina15,Shalm15}. Beyond its foundational weight, nonlocality carries a striking practical consequence: a Bell-inequality violation certifies genuinely quantum behavior from the observed statistics alone, without any modeling of the devices that produced them. This insight gave birth to the device-independent (DI) approach to quantum information~\cite{Pironio16}, in which all devices are treated as black boxes. Within this paradigm, tasks such as quantum key distribution~\cite{Ekert91,Acin07} and randomness generation~\cite{Colbeck2006,Pironio10} can be accomplished, and quantum resources themselves can be characterized, ranging from entanglement quantification~\cite{Moroder13} to the self-testing of states and measurements~\cite{Supic2020selftestingof}.
 
A related, intrinsically asymmetric form of quantum correlation is quantum steering~\cite{Schr35}: by measuring a share of an entangled state, one party updates the ensemble of conditional states held by a distant party in a manner that no local-hidden-state (LHS) model can mimic. Since its modern formulation by Wiseman, Jones, and Doherty~\cite{Wiseman07}, steering has developed into a research field of its own, with applications in one-sided DI protocols; see Refs.~\cite{Cavalcanti17,Uola2020Steering} for reviews. Its degree can be quantified by dedicated measures such as the steerable weight~\cite{SNC14} and the steering robustness~\cite{Piani15}, the latter of which enjoys an operational meaning in subchannel-discrimination tasks.
 
These two lines of research meet in the question of whether steerability can be certified, and indeed quantified, when \emph{no} device is trusted. An affirmative answer was provided by the framework of assemblage moment matrices (AMMs)~\cite{CBLC16,CBLC18}: by attaching moment matrices to the subnormalized conditional states arising in a steering experiment and matching their entries with the observed correlations, the quantification of steering robustness relaxes to a semidefinite program (SDP). This yields DI lower bounds on steerability and measurement incompatibility (see \cite{Cavalcanti16} for another approach), together with a route to self-testing~\cite{Chen2021robustselftestingof,SLChen2026universal}.
 
So far, however, these quantitative tools have been restricted to two parties. This stands in contrast with the steering literature at large, where multipartite scenarios have been formulated and explored both theoretically and experimentally~\cite{He13,Cavalcanti15}, and where they were found to harbor genuinely new physics: with two or more steering parties, there exist assemblages that obey all no-signaling-type constraints and yet admit no quantum realization. This phenomenon, known as post-quantum steering~\cite{Sainz15}, has no bipartite counterpart. On the DI side, Ref.~\cite{CBLC18} outlined the structure of AMMs in a tripartite scenario and observed that they characterize a superset of the quantum correlations which includes post-quantum steering, but did not provide any quantitative statements. A fully DI \emph{quantification} of multipartite steerability has thus been missing.
 
\begin{figure}[t]
\centering
\subfigure[]{\includegraphics[width=0.78\linewidth]{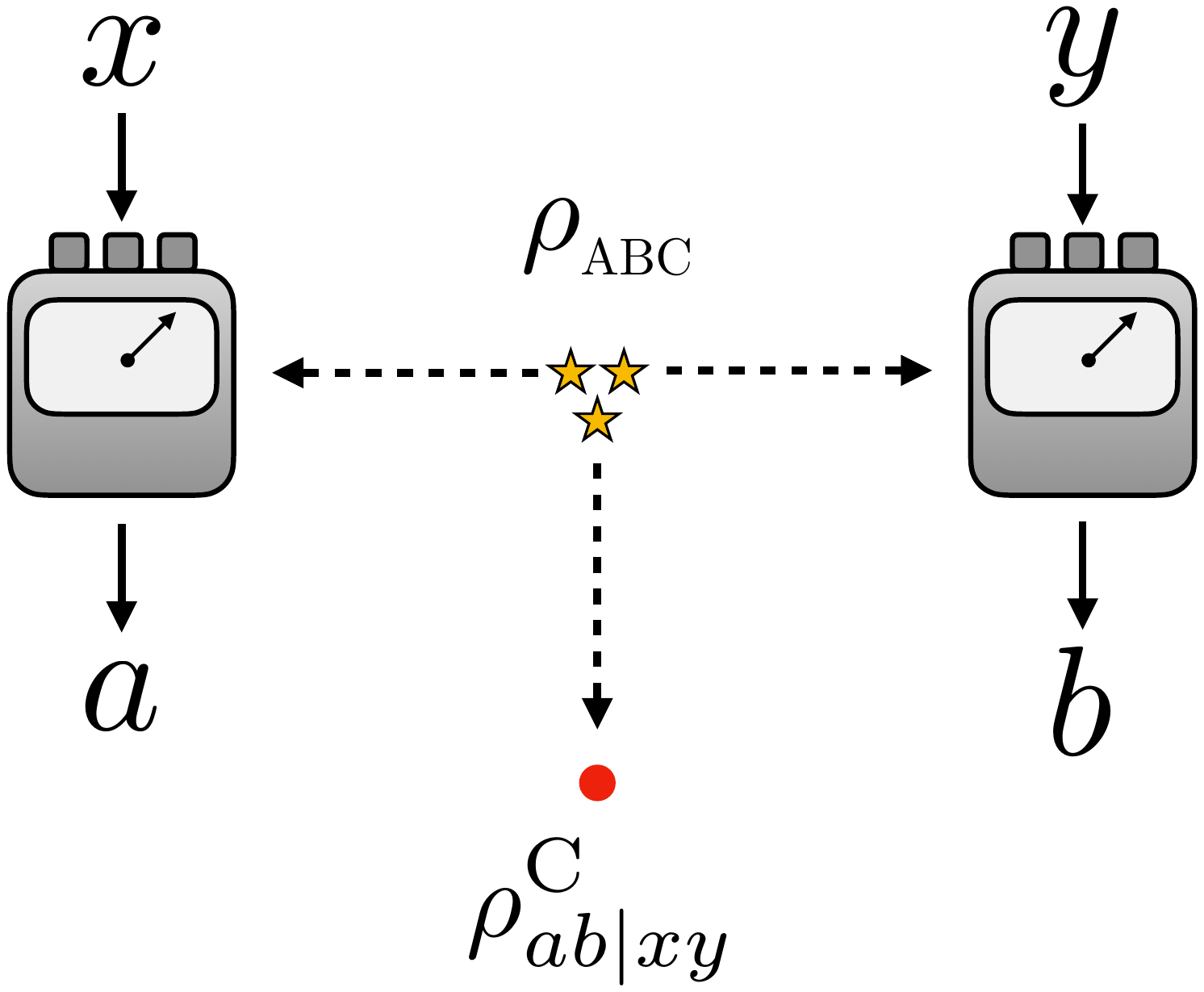}\label{Fig_scenarios_2to1}}\\
\subfigure[]{\includegraphics[width=0.78\linewidth]{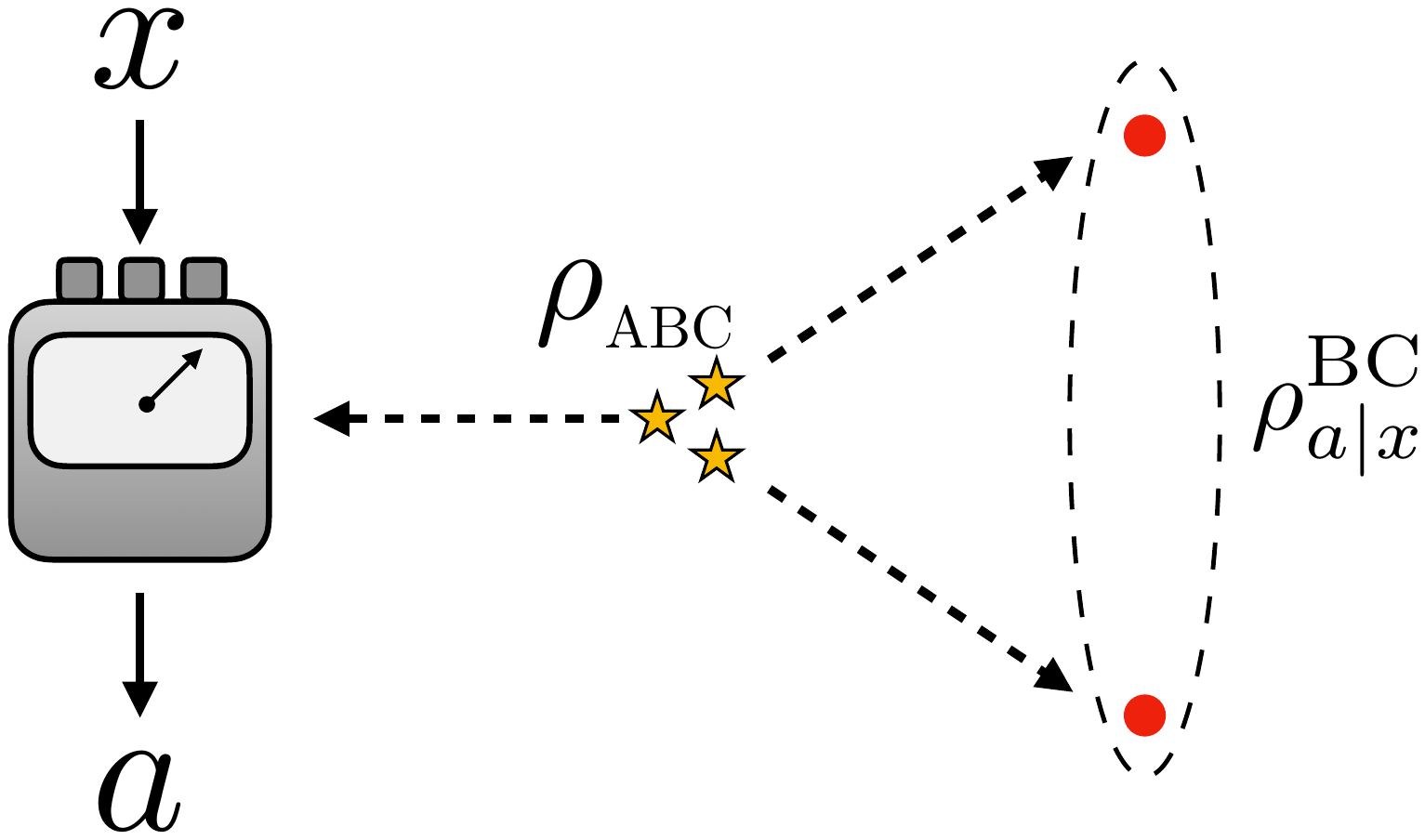}\label{Fig_scenarios_1to2}}
\caption{The two tripartite steering scenarios considered in this work. Three parties share a quantum state $\rabc$. The measurement devices of the steering parties are uncharacterized black boxes, which return an outcome for each input; the steered side is described by the conditional states. (a) The 2-steer-1 scenario: Alice and Bob, upon receiving the inputs $x$ and $y$, return the outcomes $a$ and $b$, and Charlie's conditional states are steered to the assemblage $\{\rabxy\}_{a,b,x,y}$ of \eq{Eq_assemblage_2to1}. (b) The 1-steer-2 scenario: Alice, upon receiving the input $x$, returns the outcome $a$, and Bob and Charlie are jointly left with the assemblage $\{\raxbc\}_{a,x}$ of \eq{Eq_assemblage_1to2}.}
\label{Fig_scenarios}
\end{figure}

In this work, we close this gap for the tripartite case. We develop the AMM framework for the two configurations obtained by splitting three parties into a steering and a steered side (see \fig{Fig_scenarios}): the \emph{2-steer-1} scenario, in which Alice and Bob jointly steer Charlie, and the \emph{1-steer-2} scenario, in which Alice steers the joint subsystem of Bob and Charlie. For each configuration, we construct SDPs that require only the observed correlations, or even just the observed value of a Bell functional, and deliver (i) upper bounds on the Tsirelson bound of a tripartite Bell inequality and (ii) lower bounds on the steering robustness of the underlying assemblage. Applied to the Mermin~\cite{Mermin90} and the Svetlichny~\cite{Svetlichny87} inequalities, the SDPs recover the exact Tsirelson bounds, $4$ and $4\sqrt{2}$ (up to the numerical precision), already at the first level of the hierarchy. As a by-product, we show that the observed violation also certifies, in a DI manner, the incompatibility of the measurements performed by the steering parties. For the case of two steering parties, we further derive a quantitative relation linking the steering robustness to the incompatibility of the two local measurement assemblages. In addition, we compare the two schemes from the perspective of post-quantum steering: while the 2-steer-1 characterization is inherently a superset of the quantum set, in the 1-steer-2 scenario every consistent assemblage admits a quantum realization~\cite{Gisin89,Hughston93}.

The paper is organized as follows. Section~\ref{Sec_Bell} introduces the tripartite Bell scenario together with the two Bell inequalities used throughout. Section~\ref{Sec_steering} reviews bipartite steering and defines the two tripartite steering scenarios and their steering robustness. Section~\ref{Sec_AMM} recalls the AMM construction in the bipartite case and generalizes it to both tripartite schemes. Our main results are presented in Secs.~\ref{Sec_2to1} and \ref{Sec_1to2}, for the 2-steer-1 and the 1-steer-2 scenarios, respectively. Section~\ref{Sec_conclusion} concludes with a summary and an outlook.

\section{Bell scenario}
\label{Sec_Bell}

We begin by recalling the scenario of a tripartite Bell-type experiment~\cite{Bell64,NLreview}. Three spatially separated parties, Alice, Bob, and Charlie, share a physical system distributed by some source. In each round of the experiment, they respectively choose measurement settings $x\in\{1,\dots,n_x\}$, $y\in\{1,\dots,n_y\}$, and $z\in\{1,\dots,n_z\}$, and register the corresponding measurement outcomes $a$, $b$, and $c$. When every device, from the source to the measurement apparatuses, is treated as a black box whose inner workings are unknown, the experiment is completely described by the collection of correlations $\vecP=\{P(a,b,c|x,y,z)\}$. In quantum theory, these correlations follow Born's rule, i.e.,
\begin{equation}\label{Eq_Born}
	P(a,b,c|x,y,z)=\tr\rb{\rabc\,A_{a|x}\otimes B_{b|y}\otimes C_{c|z}},
\end{equation}
where $\rabc$ is the quantum state shared by the three parties, while $\MAA$, $\MAB$, and $\{C_{c|z}\}_{z,c}$ are the positive-operator-valued measures (POVMs) describing, respectively, the local measurements of Alice, Bob, and Charlie, i.e., $A_{a|x}\succeq 0$ and $\sum_{a}A_{a|x}=\id$ for all $x$, and similarly for Bob's and Charlie's measurements.\footnote{Throughout this work, we use $A\succeq B$ to signify that $A-B$ is positive semidefinite.} We denote by $\Q$ the set of correlations admitting a quantum realization in the form of \eq{Eq_Born} for some state and some local POVMs acting on Hilbert spaces of arbitrary dimension.

On the other hand, a correlation $\vecP$ is said to be Bell local if it admits a local-hidden-variable model, i.e.,
\begin{equation}\label{Eq_LHV}
	P(a,b,c|x,y,z)=\sum_\lambda P(\lambda)P(a|x,\lambda)P(b|y,\lambda)P(c|z,\lambda)
\end{equation}
for some probability distributions $P(\lambda)$, $P(a|x,\lambda)$, $P(b|y,\lambda)$, and $P(c|z,\lambda)$. We denote the set of Bell-local correlations by $\mathcal{L}$. Correlations lying outside $\mathcal{L}$ are Bell nonlocal, and their nonlocality can be witnessed through the violation of Bell inequalities.

Throughout this work, we illustrate our results in the simplest tripartite Bell scenario, where all the inputs and outcomes are binary, i.e., $x,y,z\in\{1,2\}$ and $a,b,c\in\{+1,-1\}$. In this case, it is convenient to introduce the observables $A_x:=\sum_a a\,A_{a|x}$ (and similarly $B_y$ and $C_z$), as well as the correlators
\begin{equation}\label{Eq_correlators}
	\ew{A_xB_yC_z}:=\sum_{a,b,c}a\,b\,c\,P(a,b,c|x,y,z).
\end{equation}
For quantum correlations, \eq{Eq_Born} implies $\ew{A_xB_yC_z}=\tr\rb{\rabc\,A_x\otimes B_y\otimes C_z}$.

Two paradigmatic tripartite Bell inequalities are considered in this work. The first one is the Mermin inequality~\cite{Mermin90}
\begin{equation}\label{Eq_Mermin}
\begin{aligned}
	\M:={}&\ew{A_1B_1C_1}-\ew{A_1B_2C_2}\\
	&-\ew{A_2B_1C_2}-\ew{A_2B_2C_1}\overset{\mathcal{L}}{\leq}2,
\end{aligned}
\end{equation}
where the symbol $\mathcal{L}$ on top of the inequality sign signifies that the bound holds for all Bell-local correlations. The second one is the Svetlichny inequality~\cite{Svetlichny87}
\begin{equation}\label{Eq_Svetlichny}
\begin{aligned}
	\mathcal{S}:={}&\ew{A_1B_1C_1}+\ew{A_1B_1C_2}+\ew{A_1B_2C_1}\\
	&-\ew{A_1B_2C_2}+\ew{A_2B_1C_1}-\ew{A_2B_1C_2}\\
	&-\ew{A_2B_2C_1}-\ew{A_2B_2C_2}\overset{\mathcal{L}}{\leq}4.
\end{aligned}
\end{equation}
The Svetlichny bound of $4$ holds, in fact, not only for all Bell-local correlations, but also for hybrid models in which any two of the parties are allowed to share arbitrary nonlocal resources; a violation of \eq{Eq_Svetlichny} therefore certifies genuine tripartite nonlocality~\cite{Svetlichny87,NLreview}.

In quantum theory, the maximal attainable value of a Bell functional is often referred to as its Tsirelson bound~\cite{Tsirelson1980}. The Tsirelson bound of $\M$ is known to be $4$~\cite{Mermin90}, and it is attained by performing suitable local measurements on the Greenberger-Horne-Zeilinger (GHZ) state~\cite{GHZ89}
\begin{equation}\label{Eq_GHZ}
	\ket{{\rm GHZ}}=\frac{1}{\sqrt{2}}\rb{\ket{000}+\ket{111}},
\end{equation}
while that of $\mathcal{S}$ is known to be $4\sqrt{2}$ (see, e.g., Ref.~\cite{NLreview}). As we shall see in Secs.~\ref{Sec_2to1} and \ref{Sec_1to2}, both values can be recovered within the framework of assemblage moment matrices~\cite{CBLC16,CBLC18}.

\section{Quantum steering in bipartite and tripartite scenarios}
\label{Sec_steering}

\subsection{Bipartite scenario}
\label{Sec_steering_bipartite}

Let us now recall the notion of quantum steering~\cite{Schr35,Cavalcanti17,Uola2020Steering}, beginning with the standard bipartite scenario. Consider two parties, Alice and Bob, sharing a quantum state $\rho_{\ttAB}$, and let Alice perform one of the measurements described by the measurement assemblage~\cite{Piani15} $\MAA$. Whenever Alice performs the $x$th measurement and obtains the outcome $a$, Bob's subsystem is left in the subnormalized conditional state
\begin{equation}\label{Eq_assemblage_bipartite}
	\rax=\tr_{\tA}\rb{\rho_{\ttAB}\,A_{a|x}\otimes\id}\quad\forall\,a,x,
\end{equation}
where $P(a|x)=\tr(\rax)$ is the probability of this event and $\rax/P(a|x)$ is the corresponding normalized state. Following Ref.~\cite{Pusey13}, we refer to the collection $\SAB_{a,x}$ as a (state) assemblage. By construction, every assemblage of the form of \eq{Eq_assemblage_bipartite} satisfies the positivity and consistency conditions
\begin{equation}\label{Eq_consistency_bipartite}
	\rax\succeq0\quad\forall\,a,x,\qquad
	\sum_a\rax=\rho_{\tB}\quad\forall\,x,
\end{equation}
where $\rho_{\tB}=\tr_\tA(\rho_{\ttAB})$ is Bob's reduced state, satisfying $\tr(\rho_\tB)=1$; in particular, the sum $\sum_a\rax$ is independent of $x$, reflecting the no-signaling nature of quantum theory.

An assemblage $\SAB_{a,x}$ admits a local-hidden-state (LHS) model~\cite{Wiseman07} when there are subnormalized states $\{\sigma_\lambda\}_\lambda$, with $\sigma_\lambda\succeq0$ and $\sum_\lambda\tr(\sigma_\lambda)=1$, such that
\begin{equation}\label{Eq_LHS_bipartite}
	\rax=\sum_\lambda D(a|x,\lambda)\,\sigma_\lambda\quad\forall\,a,x,
\end{equation}
where, without loss of generality, the local response function can be taken to be deterministic, i.e., $\lambda=(\lambda_1,\dots,\lambda_{n_x})$ and $D(a|x,\lambda)=\delta_{a,\lambda_x}$~\cite{Cavalcanti17}. If such a model exists, Bob can interpret each $\rax$ as arising from a preexisting state $\sigma_\lambda$, with Alice's announcement of $(a,x)$ merely updating his classical knowledge about $\lambda$; the assemblage is then said to be unsteerable. Conversely, an assemblage that does not admit any LHS model is steerable (from Alice to Bob).

The degree of steerability of a given assemblage can be quantified, e.g., by the steering robustness ($\SR$)~\cite{Piani15}, defined as the minimal weight $t\geq0$ of another valid assemblage $\{\tau_{a|x}\}_{a,x}$ such that the mixture $\{(\rax+t\,\tau_{a|x})/(1+t)\}_{a,x}$ admits an LHS model. By absorbing the factor $(1+t)$ into the LHS members, $\SR$ can be cast as the semidefinite program (SDP)~\cite{Piani15,Cavalcanti17}
\begin{subequations}\label{Eq_SR_bipartite}
\begin{align}
	\SR(\SAB)=\min_{\{\sigma_\lambda\}}&\quad\sum_\lambda\tr(\sigma_\lambda)-1\\
	{\rm s.t.}&\quad\sum_\lambda D(a|x,\lambda)\,\sigma_\lambda\succeq\rax\quad\forall\,a,x,\\
	&\quad\sigma_\lambda\succeq0\quad\forall\,\lambda.
\end{align}
\end{subequations}
Alternative quantifiers, such as the steerable weight~\cite{SNC14}, can be treated in a completely analogous manner (see also the reviews~\cite{Cavalcanti17,Uola2020Steering}). Notably, $\SR$ enjoys a clear operational meaning: it quantifies exactly the advantage that the underlying state provides in a class of subchannel-discrimination tasks~\cite{Piani15}.

\subsection{Tripartite scenarios}
\label{Sec_steering_tripartite}

When a third party joins the experiment, inequivalent steering configurations arise, depending on which subset of the parties holds the untrusted devices and plays the role of the steering side~\cite{He13,Cavalcanti15,Sainz15}. In this work, we focus on the two scenarios associated with splitting the three parties into a steering side and a steered side, which we refer to as the 2-steer-1 scenario (Alice and Bob jointly steer Charlie) and the 1-steer-2 scenario (Alice steers Bob and Charlie), respectively. Readers can refer to Fig.~\ref{Fig_scenarios} for the two schemes.

\subsubsection{The 2-steer-1 scenario}
\label{Sec_steering_2to1}

Consider first the situation where both Alice's and Bob's devices are uncharacterized, while Charlie's subsystem is fully trusted. Whenever Alice and Bob perform the measurements labeled by $(x,y)$ and obtain the outcomes $(a,b)$, Charlie's subsystem is left in the subnormalized conditional state
\begin{equation}\label{Eq_assemblage_2to1}
	\rabxy=\tr_{\tAB}\rb{\rabc\,A_{a|x}\otimes B_{b|y}\otimes\id}\quad\forall\,a,b,x,y.
\end{equation}
In analogy with \eq{Eq_consistency_bipartite}, every assemblage $\{\rabxy\}_{a,b,x,y}$ of this form satisfies the positivity and no-signaling-type consistency conditions~\cite{Sainz15}
\begin{subequations}\label{Eq_consistency_2to1}
\begin{align}
	&\rabxy\succeq0\quad\forall\,a,b,x,y,\\
	&\sum_a\rabxy=\sum_a\rho^{\mbox{\tiny C}}_{ab|x'y}\quad\forall\,b,y,x\neq x',\\
	&\sum_b\rabxy=\sum_b\rho^{\mbox{\tiny C}}_{ab|xy'}\quad\forall\,a,x,y\neq y',\\
	&\sum_{a,b}\rabxy=\rc\quad\forall\,x,y,
\end{align}
\end{subequations}
where $\rc=\tr_{\tAB}(\rabc)$ is Charlie's reduced state, satisfying $\tr(\rc)=1$.

We can generalize the bipartite LHS model of Eq.~\eqref{Eq_LHS_bipartite} to tripartite as follows: we say that the assemblage $\{\rabxy\}_{a,b,x,y}$ admits an LHS model from Alice and Bob to Charlie when it can be decomposed as~\cite{Cavalcanti15}
\begin{equation}\label{Eq_LHS_2to1_origin}
    \rabxy=\sum_\lambda P(a,b|x,y,\lambda)\,\sigma_\lambda\quad\forall\,a,b,x,y,
\end{equation}
where $\sigma_\lambda\succeq0$, $\sum_\lambda\tr(\sigma_\lambda)=1$. Here, $\{P(a,b|x,y,\lambda)\}$ is Alice and Bob's joint response. If no decomposition of the above form exists, the assemblage is steerable from Alice and Bob to Charlie. In this work, the correlation $\{P(a,b|x,y,\lambda)\}$ will be taken in three different sets: local, quantum, and no-signaling. For the first, Eq.~\eqref{Eq_LHS_2to1_origin} becomes
\begin{equation}\label{Eq_LHS_2to1}
	\rabxy=\sum_\lambda D(a|x,\lambda)D(b|y,\lambda)\,\sigma_\lambda\quad\forall\,a,b,x,y,
\end{equation}
where, without loss of generality, the hidden variable $\lambda=(\lambda_1,\dots,\lambda_{n_x},\mu_1,\dots,\mu_{n_y})$ jointly encodes a deterministic strategy for both untrusted parties, i.e., $D(a|x,\lambda)=\delta_{a,\lambda_x}$ and $D(b|y,\lambda)=\delta_{b,\mu_y}$. The corresponding steering robustness, which we denote by $\SRL$ (``L'' for local), is defined in complete analogy with the bipartite case and can be computed via the SDP\footnote{We ignore the notation such as ``$\forall a,b,x,y$'' and ``$\forall \lambda$ when there is no risk of confusion.}
\begin{subequations}\label{Eq_SR_2to1}
\begin{align}
	\SRL(\{\rabxy\})=
    \min_{\{\sigma_\lambda\}}&\quad\sum_\lambda\tr(\sigma_\lambda)-1\\
	{\rm s.t.}&\quad\sum_\lambda D(a|x,\lambda)D(b|y,\lambda)\,\sigma_\lambda\succeq\rabxy\\
	&\quad\sigma_\lambda\succeq0.
\end{align}
\end{subequations}
Here, and in the no-signaling variant below, the noisy assemblage $\{\tau_{ab|xy}\}_{a,b,x,y}$ used for defining the robustness is chosen as one satisfying the positivity and consistency conditions of \eq{Eq_consistency_2to1}.

It is worth noting that a nonzero value of $\SRL$ does not yet demonstrate that Alice and Bob steer Charlie. To see this, consider the assemblage $\rabxy=P(a,b|x,y)\,\rho_\tC$, where Charlie holds one fixed state $\rho_\tC$ and nothing is steered at all. Taking the trace of \eq{Eq_LHS_2to1} shows that a decomposition of that form requires $P(a,b|x,y)$ to be local. Any nonlocal correlation between Alice and Bob therefore already gives $\SRL(\{\rabxy\})>0$. To attribute a nonzero value to the steering of Charlie, rather than to the correlations shared by the two steering parties, the set of responses granted to Alice and Bob has to be enlarged.

The physically natural choice is to grant them whatever quantum theory allows. Let $\mathcal{Q}$ denote the set of bipartite correlations $\{P(a,b|x,y)\}_{a,b,x,y}$ that admit a quantum realization. Requiring every response in \eq{Eq_LHS_2to1_origin} to lie in $\mathcal{Q}$ defines the steering robustness $\SRQ(\{\rabxy\})$, again as the minimal weight of an admixed assemblage that has to be added before such a model exists. A nonzero value of $\SRQ$ says that Charlie's conditional states cannot be reproduced by any preexisting quantum resource held by Alice and Bob, which is the statement one would like to certify.

This quantity, however, is not directly computable. What turns \eq{Eq_SR_2to1} into an SDP is that the local set is a polytope with finitely many extreme points: letting $\lambda$ label the $16$ deterministic strategies fixes the coefficients $D(a|x,\lambda)D(b|y,\lambda)$, so that only the hidden states $\{\sigma_\lambda\}$ remain to be optimized, and every constraint is linear in them. The same will hold for the no-signaling set below. The quantum set $\mathcal{Q}$ is convex but is not a polytope, and its extreme points form a continuum. The responses and the hidden states then both remain unknown, the products $P(a,b|x,y,\lambda)\,\sigma_\lambda$ turn the constraint bilinear, and we are not aware of any semidefinite formulation of $\SRQ$. Relaxing the quantum set $\mathcal{Q}$ to a level of the Navascu\'es-Pironio-Ac\'in hierarchy~\cite{NPA} does not remove the difficulty. Each response would then be replaced by a moment matrix, but the decomposition would still contain products of two unknowns, and $\lambda$ would still range over a continuum. A formulation in which $\lambda$ is absorbed into a separability condition, which is then relaxed in turn, may be possible, but it lies outside the scope of this work.

We therefore work with a relaxation. Since every quantum correlation is no-signaling, granting Alice and Bob an arbitrary no-signaling response can only make the model easier to satisfy, and the resulting robustness lower bounds $\SRQ$.

For $\{P(a,b|x,y,\lambda)\}$ being no-signaling (NS), Eq.~\eqref{Eq_LHS_2to1_origin} becomes
\begin{equation}\label{Eq_LHS_2to1_NS}
	\rabxy=\sum_\lambda\PNS(a,b|x,y,\lambda)\,\sigma_\lambda\quad\forall\,a,b,x,y,
\end{equation}
where each response $\{\PNS(a,b|x,y,\lambda)\}_{a,b}$ is a valid conditional distribution. For every $\lambda$, it is required to obey the no-signaling conditions
\begin{subequations}\label{Eq_NS_response}
\begin{align}
	&\sum_b\PNS(a,b|x,y,\lambda)=\sum_b\PNS(a,b|x,y',\lambda)\ \ \forall\,a,x,y\neq y',\\
	&\sum_a\PNS(a,b|x,y,\lambda)=\sum_a\PNS(a,b|x',y,\lambda)\ \ \forall\,b,y,x\neq x'.
\end{align}
\end{subequations}
In such a model, Alice and Bob may be correlated more strongly than any local strategy allows. They may even be correlated more strongly than quantum theory permits. The only requirement is that neither party can signal to the other by its choice of setting. The conditions of \eq{Eq_NS_response} are linear, and so are positivity and normalization. Together they define a polytope. Without loss of generality, we may again let $\lambda$ label its extremal points. The coefficients $\PNS(a,b|x,y,\lambda)$ are then fixed, and only the hidden states $\{\sigma_\lambda\}$ remain to be optimized. In the binary scenario, this polytope has $24$ extremal points~\cite{Barrett05PRA}. They are the $16$ deterministic strategies of \eq{Eq_LHS_2to1} and the $8$ Popescu-Rohrlich boxes~\cite{Popescu1994}. The corresponding steering robustness is therefore computed via the SDP
\begin{subequations}\label{Eq_SRNS_2to1}
\begin{align}
	\SRNS(\{\rabxy\})=
    \min_{\{\sigma_\lambda\}}&\quad\sum_\lambda\tr(\sigma_\lambda)-1\\
	{\rm s.t.}&\quad\sum_\lambda\PNS(a,b|x,y,\lambda)\,\sigma_\lambda\succeq\rabxy\\
	&\quad\sigma_\lambda\succeq0.
\end{align}
\end{subequations}
The three notions are ordered by the inclusions $\mathcal{L}\subset\mathcal{Q}\subset\mathcal{NS}$, where $\mathcal{L}$ and $\mathcal{NS}$ denote the local and the no-signaling sets of joint responses. Every feasible point of \eq{Eq_SR_2to1} is a feasible point of the model with quantum responses, which in turn is a feasible point of \eq{Eq_SRNS_2to1}, always with the same objective value. For the last step it suffices to set $\sigma_\lambda=0$ for the eight Popescu-Rohrlich boxes, the deterministic strategies being themselves extremal no-signaling boxes. We thus have
\begin{equation}\label{Eq_SRNS_vs_SR}
	\SRNS(\{\rabxy\})\leq\SRQ(\{\rabxy\})\leq\SRL(\{\rabxy\}).
\end{equation}
Certifying $\SRNS(\{\rabxy\})>0$ is therefore more demanding than certifying $\SRL(\{\rabxy\})>0$, and it is what we shall do, since by \eq{Eq_SRNS_vs_SR} it also certifies $\SRQ(\{\rabxy\})>0$.

\subsubsection{The 1-steer-2 scenario}
\label{Sec_steering_1to2}

Consider now the complementary situation, where only Alice's devices are uncharacterized and she attempts to steer the joint subsystem of Bob and Charlie. Whenever Alice performs the $x$th measurement and obtains the outcome $a$, the joint subsystem of Bob and Charlie is left in the subnormalized conditional state
\begin{equation}\label{Eq_assemblage_1to2}
	\raxbc=\tr_{\tA}\rb{\rabc\,A_{a|x}\otimes\id\otimes\id}\quad\forall\,a,x,
\end{equation}
which satisfies $\raxbc\succeq0$ for all $a,x$, as well as $\sum_a\raxbc=\rbc$ for all $x$, with $\rbc=\tr_\tA(\rabc)$ and $\tr(\rbc)=1$.

Analogously, the assemblage $\{\raxbc\}_{a,x}$ admits an LHS model from Alice to Bob and Charlie when
\begin{equation}\label{Eq_LHS_1to2}
	\raxbc=\sum_\lambda D(a|x,\lambda)\,\sigma^{\ttBC}_\lambda\quad\forall\,a,x,
\end{equation}
where $D(a|x,\lambda)$ is defined as in \eq{Eq_LHS_bipartite}, while $\{\sigma^{\ttBC}_\lambda\}_\lambda$ is a collection of subnormalized states acting on the joint Hilbert space of Bob and Charlie, with $\sigma^{\ttBC}_\lambda\succeq0$ and $\sum_\lambda\tr(\sigma^{\ttBC}_\lambda)=1$. Let us stress that the hidden states $\sigma^{\ttBC}_\lambda$ are arbitrary bipartite states; in particular, they may well be entangled across the Bob-Charlie cut. Indeed, the model of \eq{Eq_LHS_1to2} only asserts that Alice's measurement outcomes reveal nothing more than classical information $\lambda$ about a preexisting joint state of Bob and Charlie, i.e., that Alice cannot steer the pair; no separability constraint is imposed on the hidden states~\cite{Cavalcanti15}. This notion should thus be distinguished from that of genuine multipartite steering~\cite{He13}, which is beyond the scope of this work. Again, when no such decomposition exists, the assemblage is steerable from Alice to Bob and Charlie, and the associated steering robustness reads
\begin{subequations}\label{Eq_SR_1to2}
\begin{align}
	\SR(\{\raxbc\})=
    \min_{\{\sigma^{\ttBC}_\lambda\}}&\quad\sum_\lambda\tr(\sigma^{\ttBC}_\lambda)-1\\
	{\rm s.t.}&\quad\sum_\lambda D(a|x,\lambda)\,\sigma^{\ttBC}_\lambda\succeq\raxbc\quad\\
	&\quad\sigma^{\ttBC}_\lambda\succeq0\quad\forall\,\lambda.
\end{align}
\end{subequations}

Provided that the assemblage is fully known, as is the case when the steered side can perform complete state tomography, each of the SDPs above can be solved directly, yielding the corresponding degree of steerability. In a fully device-independent setting, however, no such knowledge is available: neither the assemblages nor any of the measurement operators are characterized, and one only has access to the observed correlation $\Pobs=\{\Pobsabcxyz\}$. In the next section, we recall the framework of assemblage moment matrices, which makes it possible to formulate relaxations of these SDPs directly at the level of the observed correlation.

\section{Assemblage moment matrices}
\label{Sec_AMM}

Moment matrices, i.e., matrices whose entries are expectation values of (products of) operators, play a prominent role in device-independent quantum information. The method introduced by Navascu\'es, Pironio, and Ac\'in (NPA)~\cite{NPA,NPA2008} yields an effective outer approximation of the quantum set $\Q$, while the local-level variant of Moroder \emph{et al.}~\cite{Moroder13} further enables, e.g., the device-independent quantification of entanglement. Building on these ideas, Ref.~\cite{CBLC16} put forward the assemblage-moment-matrix (AMM) framework. What sets the AMMs apart is that the matrices are attached to the \emph{subnormalized} conditional states of an assemblage, which makes them a natural tool for quantifying steerability in a device-independent fashion~\cite{CBLC16,CBLC18}. Below, we first review the AMM framework in the bipartite scenario, where it was originally formulated~\cite{CBLC16,CBLC18}, and then adapt it to the two tripartite scenarios introduced in Sec.~\ref{Sec_steering_tripartite}; the construction for the 2-steer-1 scenario was outlined in Ref.~\cite{CBLC18}, while that for the 1-steer-2 scenario is constructed here (see Ref.~\cite{Sainz15} for a related moment-matrix characterization of assemblages in multipartite steering scenarios).

\subsection{AMMs in the bipartite scenario}
\label{Sec_AMM_bipartite}

Consider again the bipartite steering scenario of Sec.~\ref{Sec_steering_bipartite}, and let $\rho$ be a (possibly subnormalized) quantum state acting on Bob's Hilbert space $\mathcal{H}_\tB$. Given a finite, ordered set $\{O_i\}_{i=1}^{m}$ of operators acting on $\mathcal{H}_\tB$, with $O_1=\id$, the associated moment matrix is defined as
\begin{equation}\label{Eq_chi_def}
	\chi[\rho]:=\sum_{i,j}\op{i}{j}\,\tr\rb{\rho\,O_j^\dagger O_i},
\end{equation}
where the dependence on the choice of $\{O_i\}_i$ is left implicit. Equivalently, $\chi[\rho]=\Lambda(\rho)$, where $\Lambda(\rho)=\sum_nK_n\rho K_n^\dagger$ is completely positive, with Kraus operators given by $K_n=\sum_i\ketbrac{i}{n}O_i$~\cite{Moroder13,CBLC16}. Two properties of $\chi[\cdot]$ are essential for what follows: (i) it is linear in $\rho$, and (ii) for every $\rho\succeq0$, one has $\chi[\rho]\succeq0$, since $\vec{v}^{\,\dagger}\chi[\rho]\vec{v}=\tr(\rho\,O_{\vec{v}}^\dagger O_{\vec{v}})\geq0$, with $O_{\vec{v}}=\sum_iv_iO_i$, for any complex vector $\vec{v}$. Moreover, since $O_1=\id$, the $(1,1)$ entry of $\chi[\rho]$ is nothing but $\tr(\rho)$; following Ref.~\cite{Moroder13}, we denote this entry by $\chi[\rho]_{\tr}$.

In the AMM framework~\cite{CBLC16}, the operators $\{O_i\}_i$ are constructed from the measurements of the steered party, which in the bipartite scenario is Bob. Specifically, the generating set is taken to be $\mathcal{G}_\tB=\{\id\}\cup\MAB$; since $\sum_bB_{b|y}=\id$ for all $y$, one may, without loss of generality, omit one outcome per setting~\cite{CBLC16}. When $\{O_i\}_i$ consists of all the products of at most $\ell$ elements of $\mathcal{G}_\tB$, the resulting moment matrix, denoted by $\chil[\rho]$, is said to be of level $\ell$; its entries then contain moments of order at most $2\ell$, and $\chil[\rho]$ is a principal submatrix of $\chi^{(\ell+1)}[\rho]$. Applying this construction to each member of the assemblage of \eq{Eq_assemblage_bipartite} gives rise to the collection $\{\chil[\rax]\}_{a,x}$, referred to as the AMMs of level $\ell$~\cite{CBLC16}. Crucially, by Born's rule and \eq{Eq_assemblage_bipartite}, the entries of $\chil[\rax]$ that are at most first order in Bob's POVM elements are directly observable in a Bell-type experiment:
\begin{equation}\label{Eq_observable_bipartite}
	\tr\rb{\rax}=P(a|x),\qquad
	\tr\rb{\rax\,B_{b|y}}=P(a,b|x,y).
\end{equation}

In the device-independent paradigm, however, neither the assemblage nor the measurement operators are characterized; only the observed correlation $\Pobs$ is available. Nevertheless, if $\Pobs$ is quantum realizable, then there must exist AMMs of every level $\ell$ whose observable entries coincide with the corresponding entries of $\Pobs$, while the remaining entries, though experimentally inaccessible, take \emph{some} consistent values. Moreover, in testing the quantum realizability of $\Pobs$, one may, without loss of generality, take all the measurements to be projective and all the unobservable moments to be real~\cite{Moroder13,CBLC18}; the former is legitimate because every POVM admits a (Naimark) dilation to a projective measurement on an enlarged Hilbert space~\cite{Peres90}, and it implies, e.g., $B_{b|y}B_{b'|y}=\delta_{b,b'}B_{b|y}$. We denote the resulting, partially characterized matrix by $\chiDI$; its entries are either fixed by $\Pobs$ or treated as real unknowns, collectively denoted by $\{u_v\}_v$.

To be explicit, consider binary inputs and outcomes, i.e., $x,y\in\{1,2\}$ and $a,b\in\{+1,-1\}$, and choose $\{O_i\}_i=\{\id,B_{+|1},B_{+|2}\}$, i.e., $\ell=1$. The projectivity of the measurements then gives, for each $(a,x)$,
\begin{equation}\label{Eq_AMM_bipartite_op}
	{\setlength{\arraycolsep}{2.5pt}
	\chi^{(1)}[\rho]=\begin{pmatrix}
	\tr(\rho) & \tr(\rho\,B_{+|1}) & \tr(\rho\,B_{+|2})\\
	\tr(\rho\,B_{+|1}) & \tr(\rho\,B_{+|1}) & \tr(\rho\,B_{+|2}B_{+|1})\\
	\tr(\rho\,B_{+|2}) & \tr(\rho\,B_{+|1}B_{+|2}) & \tr(\rho\,B_{+|2})
	\end{pmatrix}}
\end{equation}
with the shorthand $\rho\equiv\rax$, and hence~\cite{CBLC18}
\begin{equation}\label{Eq_AMMDI_bipartite}
	{\setlength{\arraycolsep}{2.5pt}
	\chiDIfirst=\begin{pmatrix}
	\pobs{a|x} & \pobs{a{+}|x1} & \pobs{a{+}|x2}\\
	\pobs{a{+}|x1} & \pobs{a{+}|x1} & u^{ax}_1\\
	\pobs{a{+}|x2} & u^{ax}_1 & \pobs{a{+}|x2}
	\end{pmatrix},}
\end{equation}
where the two experimentally inaccessible moments $\tr(\rax\,B_{+|2}B_{+|1})$ and $\tr(\rax\,B_{+|1}B_{+|2})$, being complex conjugates of each other, have been replaced by the single real unknown $u^{ax}_1$, which loses nothing since averaging any feasible point with its complex conjugate keeps it feasible with the same objective value~\cite{Moroder13}. Equipped with the device-independent AMMs, one can, e.g., outer approximate the set of quantum correlations, upper bound the Tsirelson bound of a Bell inequality, and lower bound the steering robustness of the underlying assemblage directly from $\Pobs$~\cite{CBLC16,CBLC18}; the tripartite analogs of these tasks are the subject of Secs.~\ref{Sec_2to1} and \ref{Sec_1to2}.

\subsection{AMMs for the 2-steer-1 scenario}
\label{Sec_AMM_2to1}

The generalization of the above construction to the 2-steer-1 scenario is immediate: the steered party is now Charlie, and the generating set is built from Charlie's POVM elements, i.e., $\mathcal{G}_\tC=\{\id\}\cup\{C_{c|z}\}_{z,c}$, again omitting one outcome per setting. Applying \eq{Eq_chi_def} to each member of Charlie's assemblage then gives rise to the collection of AMMs $\{\chil[\rabxy]\}_{a,b,x,y}$~\cite{CBLC18}. By Eqs.~(\ref{Eq_Born}) and (\ref{Eq_assemblage_2to1}), the observable entries now read
\begin{equation}\label{Eq_observable_2to1}
\begin{aligned}
	\tr\rb{\rabxy}&=P(a,b|x,y),\\
	\tr\rb{\rabxy\,C_{c|z}}&=P(a,b,c|x,y,z),
\end{aligned}
\end{equation}
where $P(a,b|x,y)=\sum_cP(a,b,c|x,y,z)$ denotes the Alice-Bob marginal, which is independent of $z$ owing to no-signaling. The device-independent version, $\chiDIabc$, is then obtained exactly as in the bipartite case. In particular, in the binary scenario, its first level is formally identical to \eq{Eq_AMMDI_bipartite}, with the conditioning $(a|x)$ replaced by $(ab|xy)$ and Bob's measurement operators replaced by Charlie's:
\begin{equation}\label{Eq_AMMDI_2to1}
	{\setlength{\arraycolsep}{2.5pt}
    \begin{aligned}
	&\chiDIabcfirst=\\
    &\begin{pmatrix}
	\pobs{ab|xy} & \pobs{ab{+}|xy1} & \pobs{ab{+}|xy2}\\
	\pobs{ab{+}|xy1} & \pobs{ab{+}|xy1} & u^{abxy}_1\\
	\pobs{ab{+}|xy2} & u^{abxy}_1 & \pobs{ab{+}|xy2}
	\end{pmatrix},
    \end{aligned}}
\end{equation}
where $u^{abxy}_1$ stands for the real unknown values for the inaccessible moment $\tr(\rabxy\,C_{+|2}C_{+|1})$.

\subsection{AMMs for the 1-steer-2 scenario}
\label{Sec_AMM_1to2}

In the 1-steer-2 scenario, the steered system is instead the joint subsystem of Bob and Charlie. Accordingly, the operators $\{O_i\}_i$ now act on $\mathcal{H}_\tB\otimes\mathcal{H}_\tC$, and we take the generating set to be
\begin{equation}\label{Eq_generating_1to2}
	\mathcal{G}_{\tB\tC}=\{\id\}\cup\{B_{b|y}\otimes\id\}_{y,b}\cup\{\id\otimes C_{c|z}\}_{z,c},
\end{equation}
where, again, one outcome per setting can be omitted. Since operators associated with different subsystems commute, every product of elements of $\mathcal{G}_{\tB\tC}$ can, without loss of generality, be ordered such that all of Bob's operators appear to the left of Charlie's. Applying \eq{Eq_chi_def} to each member of the assemblage of \eq{Eq_assemblage_1to2} yields the collection of AMMs $\{\chil[\raxbc]\}_{a,x}$, whose observable entries now comprise
\begin{equation}\label{Eq_observable_1to2}
\begin{aligned}
	\tr\rb{\raxbc}&=P(a|x),\\
	\tr\rb{\raxbc\,B_{b|y}\otimes\id}&=P^{\ttAB}(a,b|x,y),\\
	\tr\rb{\raxbc\,\id\otimes C_{c|z}}&=P^{\ttAC}(a,c|x,z),\\
	\tr\rb{\raxbc\,B_{b|y}\otimes C_{c|z}}&=P(a,b,c|x,y,z),
\end{aligned}
\end{equation}
where $P^{\ttAB}(a,b|x,y)$ and $P^{\ttAC}(a,c|x,z)$ denote the corresponding marginals of the tripartite distribution.

The device-independent version, $\chiDIbc$, is again obtained by fixing the observable entries to the corresponding values of $\Pobs$ and treating the remaining ones as real unknowns. In the binary scenario, taking $\ell=1$, i.e., $\{O_i\}_i=\{\id,B_{+|1}\otimes\id,B_{+|2}\otimes\id,\id\otimes C_{+|1},\id\otimes C_{+|2}\}$, one finds, for each $(a,x)$,

\begin{widetext}
\begin{equation}\label{Eq_AMMDI_1to2}
	\chiDIbcfirst=
	\begin{pmatrix}
	\pobs{a|x} & \pobsAB{a{+}|x1} & \pobsAB{a{+}|x2} & \pobsAC{a{+}|x1} & \pobsAC{a{+}|x2}\\
	\pobsAB{a{+}|x1} & \pobsAB{a{+}|x1} & u^{ax}_1 & \pobs{a{+}{+}|x11} & \pobs{a{+}{+}|x12}\\
	\pobsAB{a{+}|x2} & u^{ax}_1 & \pobsAB{a{+}|x2} & \pobs{a{+}{+}|x21} & \pobs{a{+}{+}|x22}\\
	\pobsAC{a{+}|x1} & \pobs{a{+}{+}|x11} & \pobs{a{+}{+}|x21} & \pobsAC{a{+}|x1} & u^{ax}_2\\
	\pobsAC{a{+}|x2} & \pobs{a{+}{+}|x12} & \pobs{a{+}{+}|x22} & u^{ax}_2 & \pobsAC{a{+}|x2}
	\end{pmatrix},
\end{equation}
\end{widetext}
where $u^{ax}_1$ and $u^{ax}_2$ stand for the (real) unobservable moments $\tr(\raxbc\,B_{+|2}B_{+|1}\otimes\id)$ and $\tr(\raxbc\,\id\otimes C_{+|2}C_{+|1})$, respectively, while, e.g., $\pobs{a{+}{+}|x12}$ is a shorthand for the observed tripartite probability $\pobs{a,b{=}{+},c{=}{+}|x,y{=}1,z{=}2}$. Note that $u^{ax}_1$ coincides with the unknown appearing in \eq{Eq_AMMDI_bipartite}, since $\tr(\raxbc\,B_{+|2}B_{+|1}\otimes\id)=\tr(\rax\,B_{+|2}B_{+|1})$ with $\rax=\tr_\tC(\raxbc)$.

We conclude this section with two remarks. First, for any legitimate assemblage, the AMMs of every level are positive semidefinite and, by linearity, the consistency conditions satisfied by the assemblage [cf.~\eq{Eq_consistency_2to1} and the discussion below \eq{Eq_assemblage_1to2}] descend to linear constraints among the AMMs; for instance, $\sum_{a,b}\chiDIabc$ must be independent of $(x,y)$, while $\sum_{a}\chiDIbc$ must be independent of $x$. These constraints, together with the positive semidefiniteness of the device-independent AMMs and the matching with the observed data, constitute the basic ingredients of the SDP relaxations formulated in Secs.~\ref{Sec_2to1} and \ref{Sec_1to2}. Second, increasing the level $\ell$ can only tighten the resulting device-independent characterizations since $\chil[\cdot]$ is a principal submatrix of $\chi^{(\ell+1)}[\cdot]$.

\section{Device-independent quantification in the 2-steer-1 scenario}
\label{Sec_2to1}

We are now in a position to quantify, in a fully device-independent manner, the steerability in the 2-steer-1 scenario [see \fig{Fig_scenarios}(a)]. Specifically, given only the observed correlation $\Pobs$, or even just the observed value of a Bell functional, we show how the AMMs of Sec.~\ref{Sec_AMM_2to1} allow one to (i) upper bound the Tsirelson bound of any tripartite Bell inequality and (ii) lower bound the steering robustness of the underlying assemblage $\{\rabxy\}_{a,b,x,y}$. Throughout this section, we write a generic Bell functional as
\begin{equation}\label{Eq_Bell_functional}
	\mathcal{B}:=\sum_{a,b,c,x,y,z}\beta^{xyz}_{abc}\,P(a,b,c|x,y,z),
\end{equation}
with real coefficients $\beta^{xyz}_{abc}$; the Mermin and the Svetlichny functionals of Eqs.~(\ref{Eq_Mermin}) and (\ref{Eq_Svetlichny}) are recovered with suitable choices of $\beta^{xyz}_{abc}$.

\subsection{Upper bounds on Tsirelson bounds}
\label{Sec_Tsirelson_2to1}

As discussed in Sec.~\ref{Sec_AMM}, if a correlation $\vecP$ is quantum realizable, then there exist AMMs of every level whose observable entries coincide with the entries of $\vecP$, which are positive semidefinite, and which inherit the consistency conditions of \eq{Eq_consistency_2to1}. Consequently, maximizing the Bell functional $\mathcal{B}$ over all matrices compatible with these requirements can only overestimate the maximum of $\mathcal{B}$ over the quantum set $\Q$. In other words, for any level $\ell$, the optimum of the SDP
\begin{subequations}\label{Eq_Tsirelson_2to1}
\begin{align}
	\mathcal{B}^{(\ell)}_{\max}&:=\\
    \max_{\vecP,\,\{u_v\}}&\quad\mathcal{B}\label{Eq_Tsirelson_2to1_obj}\\
	{\rm s.t.}\ \ &\quad\chiDIabc\succeq0\quad\forall\,a,b,x,y,\label{Eq_Tsirelson_2to1_psd}\\
	&\quad\sum_a\chiDIabc=\sum_a\chilbb[\rho^{\mbox{\tiny C}}_{ab|x'y}]\label{Eq_Tsirelson_2to1_nsx}\\
	&\hspace{11em}\forall\,b,y,x\neq x',\nonumber\\
	&\quad\sum_b\chiDIabc=\sum_b\chilbb[\rho^{\mbox{\tiny C}}_{ab|xy'}]\label{Eq_Tsirelson_2to1_nsy}\\
	&\hspace{11em}\forall\,a,x,y\neq y',\nonumber\\
	&\quad\sum_{a,b}\chiDIabc_{\tr}=1\label{Eq_Tsirelson_2to1_norm}
\end{align}
\end{subequations}
is an upper bound on the Tsirelson bound of $\mathcal{B}$. Here, no entry is fixed by data: the optimization variables comprise both the probabilities $\vecP$, which enter the matrices $\{\chiDIabc\}_{a,b,x,y}$ through the identification of Sec.~\ref{Sec_AMM_2to1}, and the unknowns $\{u_v\}_v$.

Applying \eq{Eq_Tsirelson_2to1} to the Mermin and the Svetlichny inequalities, we find that, already at the first level, the resulting upper bounds coincide, within the numerical precision, with the known Tsirelson bounds, namely, $4$ and $4\sqrt{2}$, respectively. The results are summarized in Table~\ref{Table_Tsirelson}. Since these values are attained by performing suitable local measurements on the GHZ state (cf.\ Sec.~\ref{Sec_results_2to1}), the AMM characterization thus recovers the Tsirelson bounds of both inequalities.

\begin{table}[t]
\caption{Upper bounds on the Tsirelson bounds of the Mermin and the Svetlichny inequalities, computed from the SDP of \eq{Eq_Tsirelson_2to1} at level $\ell=1$. In both cases, the first-level bound already coincides with the known Tsirelson bound within the numerical precision of $10^{-6}$.}
\label{Table_Tsirelson}
\begin{ruledtabular}
\begin{tabular}{lccc}
	Bell inequality & Local & AMM bound & Tsirelson\\
	 & bound & ($\ell=1$) & bound\\
	\hline
	Mermin, \eq{Eq_Mermin} & $2$ & $4.0000$ & $4$\\
	Svetlichny, \eq{Eq_Svetlichny} & $4$ & $5.6569$ & $4\sqrt{2}$\\
\end{tabular}
\end{ruledtabular}
\end{table}

Let us remark that, in the tripartite scenario, the characterization provided by the AMMs is not expected to converge to the quantum set even in the limit $\ell\to\infty$: there exist assemblages that satisfy all the consistency conditions of \eq{Eq_consistency_2to1}, and hence give rise to positive semidefinite AMMs of every level, yet admit no quantum realization, a phenomenon known as post-quantum steering~\cite{Sainz15}; see also the discussion in Ref.~\cite{CBLC18}. This, however, does not compromise the results presented here: the SDPs of this section are relaxations, so their optima remain valid upper bounds on Tsirelson bounds. As we have just seen, for the two inequalities considered, the upper bounds are in fact tight.

\subsection{Lower bounds on steering robustness}
\label{Sec_SRDI_2to1}

\subsubsection{Local joint response}
\label{Sec_results_2to1}

\begin{figure}[t]
\centering
\subfigure[]{\includegraphics[width=0.85\linewidth]{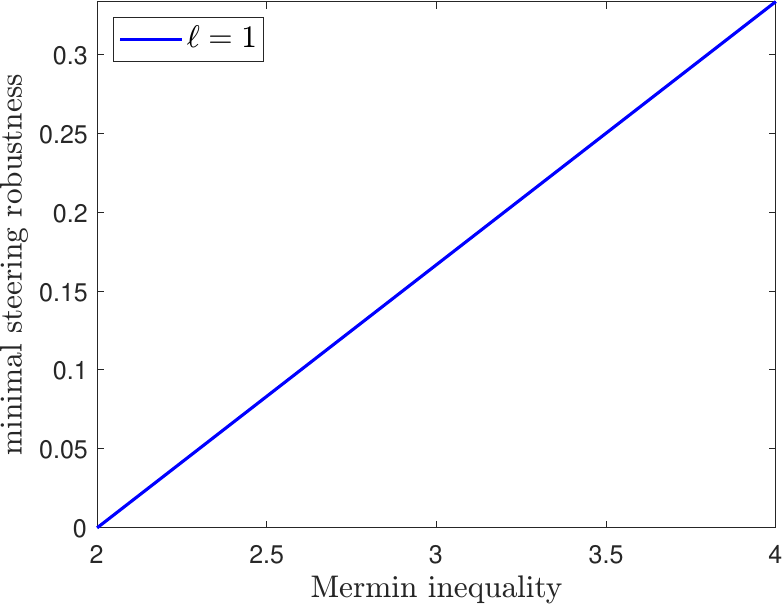}\label{Fig_SR_2to1_Mermin}}\\
\subfigure[]{\includegraphics[width=0.85\linewidth]{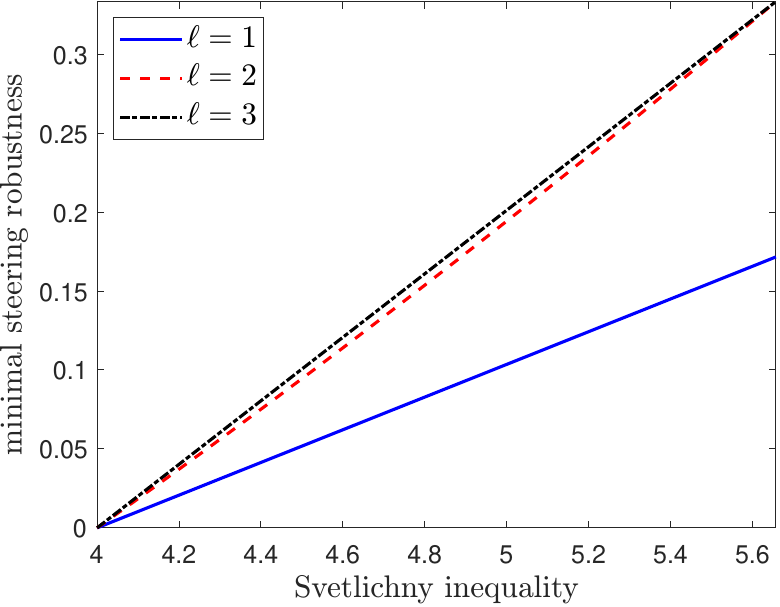}\label{Fig_SR_2to1_Svet}}
\caption{Device-independent lower bounds on the steering robustness $\SRL$ in the 2-steer-1 scenario, obtained from the SDP of \eq{Eq_SRDI_2to1}, as functions of the observed violation of (a) the Mermin inequality and (b) the Svetlichny inequality. Here and in the figures below, the hierarchy is truncated once the next level improves the optimum by less than $10^{-7}$, and every level computed is plotted, so the highest one shown is the converged bound.}
\label{Fig_SR_2to1}
\end{figure}

We now turn to the device-independent quantification of steerability. The key observation~\cite{CBLC16} is that the map $\Lambda$ underlying the AMM construction, being linear and completely positive, preserves the operator order: $A\succeq B$ implies $\Lambda(A)=\Lambda(B)+\Lambda(A-B)\succeq\Lambda(B)$. Applying $\Lambda$ to both sides of each constraint of SDP of $\SRL$ (\eq{Eq_SR_2to1}) therefore shows that, whenever $\{\sigma_\lambda\}_\lambda$ is feasible for \eq{Eq_SR_2to1},
\begin{equation}\label{Eq_lifted_LHS_2to1}
	\sum_\lambda D(a|x,\lambda)D(b|y,\lambda)\,\chil[\sigma_\lambda]\succeq\chil[\rabxy]
\end{equation}
for all $a,b,x,y$, while $\chil[\sigma_\lambda]\succeq0$ and $\chil[\sigma_\lambda]_{\tr}=\tr(\sigma_\lambda)$. Every feasible point of \eq{Eq_SR_2to1} is thus mapped to a collection of moment matrices satisfying the analogous constraints, \emph{with the same objective value}. Relaxing the latter to the device-independent matrices of Sec.~\ref{Sec_AMM_2to1} then yields a lower bound on $\SRL(\{\rabxy\})$. Explicitly, given the observed value $\Bobs$ of a Bell functional $\mathcal{B}$, we solve
\begin{subequations}\label{Eq_SRDI_2to1}
\begin{align}
	\SRDItwoone&(\Bobs)=\\
    \min_{\{u_v\}} \ &\quad\sum_\lambda\chilbb[\sigma_\lambda]_{\tr}-1\label{Eq_SRDI_2to1_obj}\\
	{\rm s.t.}\ \ &\quad\sum_\lambda D(a|x,\lambda)D(b|y,\lambda)\,\chilbb[\sigma_\lambda]\succeq\chiDIabc\label{Eq_SRDI_2to1_lhs}\\
	&\quad\chilbb[\sigma_\lambda]\succeq0\quad\forall\,\lambda,\label{Eq_SRDI_2to1_psd}\\
	&\quad\text{Eqs.~(\ref{Eq_Tsirelson_2to1_psd})--(\ref{Eq_Tsirelson_2to1_norm})},\label{Eq_SRDI_2to1_cons}\\
	&\quad\mathcal{B}=\Bobs,\label{Eq_SRDI_2to1_data}
\end{align}
\end{subequations}
where each $\chilbb[\sigma_\lambda]$ is a matrix built from the same operator set as $\chiDIabc$, and hence subject to the same projectivity-induced relations among its entries. The difference is that all the entries of $\chilbb[\sigma_\lambda]$ are treated as unknowns, and the optimization runs over all the free entries of the matrices involved. For any quantum realization compatible with the observed data, we then have
\begin{equation}\label{Eq_SR_chain_2to1}
	\SRL(\{\rabxy\})\geq\SRDItwoone(\Pobs)\geq\SRDItwoone(\Bobs),
\end{equation}
where $\SRDItwoone(\Pobs)$ denotes the optimum obtained when the constraint of \eq{Eq_SRDI_2to1_data} is replaced by the full data-matching condition $P(a,b,c|x,y,z)=\Pobsabcxyz$ for all $a,b,c,x,y,z$. The first inequality holds because the SDP is a relaxation of \eq{Eq_SR_2to1}, and the second because fixing only $\Bobs$ retains strictly less information than fixing $\Pobs$. While the latter version yields bounds that are at least as tight, the former has the practical advantage of requiring the estimation of a single number only.

We have solved the SDP of \eq{Eq_SRDI_2to1} for the Mermin and the Svetlichny functionals, varying $\Bobs$ between the corresponding local bound and Tsirelson bound. All SDPs are solved with standard semidefinite-programming packages. The results are shown in \fig{Fig_SR_2to1}. In all cases, the optimal bounds vanish at the local bound and increase linearly with $\Bobs$, reaching their maximal values at the corresponding Tsirelson bounds; the former is expected, since Bell-local correlations certify no steering.
In every case we increase $\ell$ until the next level improves the optimum by less than $10^{-7}$, and we plot all the levels computed in this way, so that the highest one shown is the converged value. For the Mermin functional the first level is already converged. For the Svetlichny functional it is not: $\ell=1$ is visibly weaker, and convergence is reached at $\ell=3$, which nearly coincides with $\ell=2$.

\subsubsection{No-signaling joint response}
\label{Sec_SRDI_2to1_NS}

\begin{figure}[t]
\centering
\includegraphics[width=0.85\linewidth]{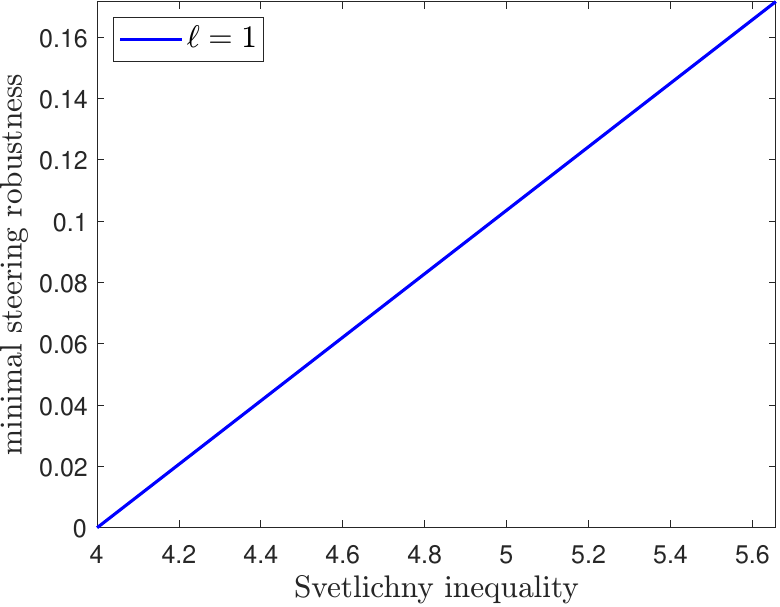}
\caption{Device-independent lower bound on the no-signaling steering robustness $\SRNS$ in the 2-steer-1 scenario, as a function of the observed violation of the Svetlichny inequality. The bound vanishes at the Svetlichny bound $\Bobs=4$ rather than at a smaller value, for the reason given below \eq{Eq_Svet_reduced}. The corresponding bound for the Mermin inequality is identically zero and is not shown.}
\label{Fig_SRNS_2to1_Svet}
\end{figure}

The same lifting applies to the no-signaling model. Replacing the local response $D(a|x,\lambda)D(b|y,\lambda)$ of \eq{Eq_SR_2to1} by $\PNS(a,b|x,y,\lambda)$ and repeating the argument of Sec.~\ref{Sec_results_2to1} gives
\begin{subequations}\label{Eq_SRDI_2to1_NS}
\begin{align}
	\SRDINStwoone&(\Bobs)=\\
    \min_{\{u_v\}} \ &\quad\sum_\lambda\chilbb[\sigma_\lambda]_{\tr}-1\label{Eq_SRDI_2to1_NS_obj}\\
	{\rm s.t.}\ \ &\quad\sum_\lambda\PNS(a,b|x,y,\lambda)\,\chilbb[\sigma_\lambda]\succeq\chiDIabc\label{Eq_SRDI_2to1_NS_lhs}\\
	&\quad\chilbb[\sigma_\lambda]\succeq0\quad\forall\,\lambda,\label{Eq_SRDI_2to1_NS_psd}\\
	&\quad\text{Eqs.~(\ref{Eq_Tsirelson_2to1_psd})--(\ref{Eq_Tsirelson_2to1_norm})},\label{Eq_SRDI_2to1_NS_cons}\\
	&\quad\mathcal{B}=\Bobs,\label{Eq_SRDI_2to1_NS_data}
\end{align}
\end{subequations}
where $\lambda$ now runs over the $24$ extremal no-signaling boxes listed below \eq{Eq_NS_response}. In parallel with \eq{Eq_SR_chain_2to1}, any quantum realization compatible with the observed data satisfies
\begin{equation}\label{Eq_SR_chain_2to1_NS}
	\SRNS(\{\rabxy\})\geq\SRDINStwoone(\Pobs)\geq\SRDINStwoone(\Bobs).
\end{equation}

We have solved \eq{Eq_SRDI_2to1_NS} for the same two functionals. They now behave very differently. For the Svetlichny inequality the bound stays nonzero, as shown in \fig{Fig_SRNS_2to1_Svet}. As can be seen, the bound vanishes at $\Bobs=4$ and grows linearly, reaching $3-2\sqrt{2}$ at the Tsirelson bound $4\sqrt{2}$. The bounds between agree with the following form
\begin{equation}\label{Eq_SRNS_DI_Svet}
	\SRDINStwoone(\Bobs)=\frac{\sqrt{2}-1}{4}\,\big(\Bobs-4\big),
\end{equation}
up to a numerical precision of $10^{-7}$. Enlarging the hidden-variable models from local to no-signaling therefore costs about half of the certifiable steering robustness at maximal violation. For the Mermin inequality the bound is \emph{zero} over the whole range $2\leq\Bobs\leq4$.

The vanishing bound is due to the structure of the no-signaling model. Combining \eq{Eq_LHS_2to1_NS} with the Born rule gives, for any assemblage admitting such a model,
\begin{equation}\label{Eq_NS_hybrid_correlation}
	P(a,b,c|x,y,z)=\sum_\lambda q_\lambda\,\PNS(a,b|x,y,\lambda)\,P(c|z,\lambda),
\end{equation}
with $q_\lambda=\tr(\sigma_\lambda)$ and $P(c|z,\lambda)=\tr(C_{c|z}\sigma_\lambda)/q_\lambda$. Alice and Bob may thus share an arbitrary no-signaling resource, while Charlie is correlated with them only through $\lambda$. Since the Bell functionals are linear and every single-party response is a mixture of deterministic ones, it suffices to examine \eq{Eq_NS_hybrid_correlation} for deterministic $C_1=c_1$ and $C_2=c_2$ with $c_1,c_2\in\{\pm1\}$.

Consider the Mermin functional of \eq{Eq_Mermin} under this substitution. It becomes
\begin{equation}\label{Eq_Mermin_reduced}
	c_1\big(\ew{A_1B_1}-\ew{A_2B_2}\big)-c_2\big(\ew{A_1B_2}+\ew{A_2B_1}\big),
\end{equation}
which for $c_1=-c_2=1$ is a CHSH functional for Alice and Bob. A Popescu-Rohrlich box attains its algebraic maximum $4$. The Tsirelson bound of the Mermin inequality is also $4$, as listed in Table~\ref{Table_Tsirelson}. Every Mermin value that quantum theory can produce is therefore already reproduced by an assemblage with a no-signaling model, and the bound vanishes identically. Zero is here the exact optimum of the problem that \eq{Eq_SRDI_2to1_NS} relaxes, so no higher level can improve it.

The same substitution in the Svetlichny functional of \eq{Eq_Svetlichny} gives
\begin{equation}\label{Eq_Svet_reduced}
	(c_1+c_2)\big(\ew{A_1B_1}-\ew{A_2B_2}\big)+(c_1-c_2)\big(\ew{A_1B_2}+\ew{A_2B_1}\big).
\end{equation}
One of the two prefactors always vanishes, so \eq{Eq_Svet_reduced} equals $\pm2(\ew{A_1B_1}-\ew{A_2B_2})$ or $\pm2(\ew{A_1B_2}+\ew{A_2B_1})$. Each is bounded by $4$ even for arbitrary no-signaling boxes. This reproduces the Svetlichny bound, which for that inequality coincides with the local bound. A violation beyond $4$ cannot be produced in this way, which is why the bound of \fig{Fig_SRNS_2to1_Svet} is nonzero precisely above $\Bobs=4$.

\begin{table*}[t]
\caption{Quantum strategies that attain the two end points of the line of \eq{Eq_SRNS_DI_Svet} in \fig{Fig_SRNS_2to1_Svet}. At the lower end point every outcome is deterministic and the induced assemblage $\rabxy=P(a,b|x,y)\ket{0}\bra{0}$ admits a local, hence no-signaling, model. At the upper end point, solving \eq{Eq_SRNS_2to1} for the induced assemblage, with $\lambda$ running over the $24$ extremal no-signaling boxes, gives $0.1715729$, in agreement with $3-2\sqrt{2}$ to seven decimal places.}
\label{Table_endpoints_2to1}
\begin{ruledtabular}
\begin{tabular}{lccccc}
End point & State & $(A_1,A_2)$ & $(B_1,B_2)$ & $(C_1,C_2)$ & $\SRNS(\{\rabxy\})$\\
\hline
$\mathcal{S}=4$ & $\ket{000}$ & $(\sigma_z,\sigma_z)$ & $(\sigma_z,\sigma_z)$ & $(\sigma_z,-\sigma_z)$ & $0$\\
$\mathcal{S}=4\sqrt{2}$ & $\ket{\rm GHZ}$ & $(\sigma_x,\sigma_y)$ & $(\sigma_x,\sigma_y)$ & $\big(\tfrac{\sigma_x-\sigma_y}{\sqrt{2}},\tfrac{\sigma_x+\sigma_y}{\sqrt{2}}\big)$ & $3-2\sqrt{2}$\\
\end{tabular}
\end{ruledtabular}
\end{table*}

Regarding the tightness of the bound, Table~\ref{Table_endpoints_2to1} lists a quantum strategy for each end point of the line. Using the argument of convexity, every point lying on the line can be reproduced by the convex mixture of the two quantum strategies. Such a mixture cannot fall below the line either, since \eq{Eq_SRNS_DI_Svet} is a lower bound on $\SRNS$. The bound is therefore tight on the whole range $4\leq\Bobs\leq4\sqrt{2}$.

We note that, in contrast with the local case, a nonzero value of $\SRNS$ cannot be produced by nonlocality between Alice and Bob alone, since the correlations between Alice and Bob already grants them an arbitrary no-signaling resource (i.e., \eq{Eq_LHS_2to1_NS}).

\subsection{Implications for measurement incompatibility}
\label{Sec_IR_2to1}

Finally, we show that the device-independent bounds obtained above also certify the incompatibility of the measurements performed by the steering parties. Recall that a measurement assemblage $\{M_{m|w}\}_{w,m}$ is jointly measurable if there exists a single ``parent'' POVM $\{G_\gamma\}_\gamma$ such that $M_{m|w}=\sum_\gamma D(m|w,\gamma)\,G_\gamma$ for all $w,m$, and that its incompatibility robustness $\IR(\{M_{m|w}\})$ is the minimal $t\geq0$ for which one can find a measurement assemblage $\{N_{m|w}\}_{w,m}$ rendering the mixture $\{(M_{m|w}+t\,N_{m|w})/(1+t)\}_{w,m}$ jointly measurable~\cite{Uola15,Designolle19} (see Ref.~\cite{Guhne2023} for a review). The use of incompatible measurements is necessary for demonstrating steering~\cite{Quint14,Uola14,Uola15}. In the bipartite scenario this is made quantitative by $\IR(\MAA)\geq\SR(\SAB_{a,x})$~\cite{CBLC16,Cavalcanti16}. In the 2-steer-1 scenario, we prove the following analog.

\begin{lemma}\label{Lem_IR_2to1}
Let $\{\rabxy\}_{a,b,x,y}$ be an assemblage of the form of \eq{Eq_assemblage_2to1}, generated by the measurement assemblages $\MAA$ and $\MAB$ acting on a tripartite state $\rabc$. Then
\begin{equation}\label{Eq_IR_2to1}
    \begin{aligned}
        &\Big[1+\max\{\IR(\MAA),\IR(\MAB)\}\Big]^2\geq\\
        &[1+\IR(\MAA)][1+\IR(\MAB)]\geq\\
        &1+\SRL(\{\rabxy\}).
    \end{aligned}
\end{equation}
\end{lemma}
The proof is given in Appendix~\ref{Sec_App_IR_2to1}. The middle expression of \eq{Eq_IR_2to1} is the one that comes out of the construction. The first inequality weakens it into a statement about a single party, which is the form used in \eq{Eq_IR_DI_2to1} below.

In particular, if both $\MAA$ and $\MAB$ are jointly measurable, the assemblage is unsteerable. Combining Eqs.~(\ref{Eq_SR_chain_2to1}) and (\ref{Eq_IR_2to1}), our device-independent bounds thus certify, from the observed value $\Bobs$ alone, a minimal degree of incompatibility of the measurements performed by (at least one of) the steering parties:
\begin{equation}\label{Eq_IR_DI_2to1}
\begin{split}
	\max\{\IR(\MAA),\IR(\MAB)\}\\
	\geq\sqrt{1+\SRDItwoone(\Bobs)}-1.
\end{split}
\end{equation}

Everything above refers to the LHS model with Alice and Bob sharing local correlations (i.e., \eq{Eq_LHS_2to1}). We now explore what happens for the LHS model with Alice and Bob sharing no-signaling correlations (i.e., \eq{Eq_LHS_2to1_NS}).

From \eq{Eq_SRNS_vs_SR}, we have $\SRNS(\{\rabxy\})\leq\SRL(\{\rabxy\})$. Lemma~\ref{Lem_IR_2to1} therefore yields
\begin{equation}\label{Eq_IR_SRNS_weak}
    \begin{aligned}
        &[1+\IR(\MAA)][1+\IR(\MAB)]\\
        &\qquad\geq1+\SRNS(\{\rabxy\}).
    \end{aligned}
\end{equation}
Observing $\SRNS(\{\rabxy\})>0$ therefore still certifies that at least one of the two steering parties performs incompatible measurements. One may expect \eq{Eq_IR_SRNS_weak} to be loose, since it was derived for a smaller set of hidden-variable models and nothing in it refers to the no-signaling structure. Proposition~\ref{Prop_IRNS_product} below shows that a notion of incompatibility adapted to that structure returns exactly \eq{Eq_IR_SRNS_weak}.

The no-signaling model calls for a notion of incompatibility of its own. In that model, Alice and Bob act as a single steering party with joint input $(x,y)$ and joint output $(a,b)$. The relevant object is then the joint measurement assemblage $\MAAB$. Joint measurability is replaced by the following notion.

\begin{definition}\label{Def_NS_simulable}
A measurement assemblage $\{M_{ab|xy}\}_{x,y,a,b}$ is \emph{no-signaling simulable} if there exist a single parent POVM $\{G_\lambda\}_\lambda$ and, for each $\lambda$, a no-signaling box $V_\lambda(a,b|x,y)$ obeying the conditions of \eq{Eq_NS_response} such that
\begin{equation}\label{Eq_NS_simulable}
	M_{ab|xy}=\sum_\lambda V_\lambda(a,b|x,y)\,G_\lambda\quad\forall\,a,b,x,y.
\end{equation}
Its no-signaling simulability robustness $\IRNS(\{M_{ab|xy}\})$ is the smallest $t\geq0$ for which some measurement assemblage $\{N_{ab|xy}\}_{x,y,a,b}$ renders the mixture $\{(M_{ab|xy}+t\,N_{ab|xy})/(1+t)\}_{x,y,a,b}$ no-signaling simulable.
\end{definition}

In \eq{Eq_NS_simulable}, $\{G_\lambda\}_\lambda$ is an arbitrary POVM on Alice and Bob's joint space. The noise $\{N_{ab|xy}\}_{x,y,a,b}$ is also an arbitrary joint measurement assemblage. 

Definition~\ref{Def_NS_simulable} is a variant of the measurement simulability framework of Ref.~\cite{Guerini2017}, where a set of measurements is reproduced by classically processing the outcome of an accessible measurement. The only change is that the post-processings are enlarged from classical responses to no-signaling boxes. The resulting class sits between two familiar ones. Restricting each $V_\lambda$ to a product of single-party deterministic strategies gives the local case of \eq{Eq_NS_simulable_JM} below. In the other direction, every no-signaling box is a mixture of deterministic responses $D(a,b|x,y,\lambda)$, so no-signaling simulability implies joint measurability of the family $\MAAB$. Note that a deterministic response of that form need not itself be no-signaling, since it may let $a$ depend on $y$.

The notion is consistent with the local case. If $\MAA$ and $\MAB$ are separately jointly measurable, say $A_{a|x}=\sum_\mu D(a|x,\mu)G^{\tA}_\mu$ and $B_{b|y}=\sum_\nu D(b|y,\nu)G^{\tB}_\nu$, then
\begin{equation}\label{Eq_NS_simulable_JM}
	A_{a|x}\otimes B_{b|y}=\sum_{\mu,\nu}D(a|x,\mu)D(b|y,\nu)\,G^{\tA}_\mu\otimes G^{\tB}_\nu.
\end{equation}
The box $D(a|x,\mu)D(b|y,\nu)$ is local, hence no-signaling, and $\{G^{\tA}_\mu\otimes G^{\tB}_\nu\}_{\mu,\nu}$ is a POVM on Alice and Bob's joint space. Thus $\IRNS(\MAAB)=0$ in that case.

\begin{proposition}\label{Prop_IRNS_product}
Let $\MAA$ and $\MAB$ act on spaces of dimension $d_\tA$ and $d_\tB$. Then
\begin{equation}\label{Eq_IRNS_product_main}
	\begin{aligned}
	&1+\IRNS(\MAAB)\\
	&\qquad=\big[1+\IR(\MAA)\big]\big[1+\IR(\MAB)\big].
	\end{aligned}
\end{equation}
\end{proposition}
The proof is given in Appendix~\ref{Sec_App_product}.

\begin{corollary}\label{Cor_IRNS_2to1}
Let $\{\rabxy\}_{a,b,x,y}$ be an assemblage of the form of \eq{Eq_assemblage_2to1}, generated by $\MAA$ and $\MAB$ acting on a tripartite state $\rabc$. Then
\begin{equation}\label{Eq_IRNS_2to1}
	\begin{aligned}
	&\IRNS(\MAAB)\\
	&\qquad\geq\SRL(\{\rabxy\})\geq\SRNS(\{\rabxy\}).
	\end{aligned}
\end{equation}
\end{corollary}

The proof follows from Proposition~\ref{Prop_IRNS_product} and Lemma~\ref{Lem_IR_2to1}, as shown below. A direct proof of the outer inequality, which does not rely on Proposition~\ref{Prop_IRNS_product}, is given in Appendix~\ref{Sec_App_NS}. Indeed, Proposition~\ref{Prop_IRNS_product} turns the middle expression of \eq{Eq_IR_2to1} into $1+\IRNS(\MAAB)$, so Lemma~\ref{Lem_IR_2to1} reads $1+\IRNS(\MAAB)\geq1+\SRL(\{\rabxy\})$. The second inequality of \eq{Eq_IRNS_2to1} is from \eq{Eq_SRNS_vs_SR}. Note that the first inequality bounds the larger quantity $\SRL$, not only $\SRNS$. Besides, Corollary~\ref{Cor_IRNS_2to1} settles the question raised below \eq{Eq_IR_SRNS_weak}. The no-signaling model suggests its own certification, $\IRNS(\MAAB)\geq\SRNS(\{\rabxy\})$, which is the outer inequality of \eq{Eq_IRNS_2to1} and is also proved directly in Appendix~\ref{Sec_App_NS}. By \eq{Eq_IRNS_product_main}, its left-hand side equals $[1+\IR(\MAA)][1+\IR(\MAB)]-1$, so this certification is \eq{Eq_IR_SRNS_weak} again. 

\section{Device-independent quantification in the 1-steer-2 scenario}
\label{Sec_1to2}

We now carry out the analogous program for the 1-steer-2 scenario [see \fig{Fig_scenarios}(b)], in which Alice alone steers the joint subsystem of Bob and Charlie. The logic parallels that of Sec.~\ref{Sec_2to1}, with the AMMs of Sec.~\ref{Sec_AMM_2to1} replaced by those of Sec.~\ref{Sec_AMM_1to2}; we therefore keep the presentation brief and highlight the differences. Two features are worth emphasizing from the outset. First, the matrices $\chiDIbc$ are labeled by $(a,x)$ only, and each of them already contains the full tripartite correlation among its entries [cf.~\eq{Eq_AMMDI_1to2}]. Second, the hidden-state matrices entering the relaxed LHS constraint below are built from the joint operator set of \eq{Eq_generating_1to2}; in particular, they accommodate arbitrary correlations between Bob and Charlie, including entanglement, in accordance with the LHS model of \eq{Eq_LHS_1to2}.

\subsection{Upper bounds on Tsirelson bounds}
\label{Sec_Tsirelson_1to2}

In complete analogy with \eq{Eq_Tsirelson_2to1}, an upper bound on the Tsirelson bound of a Bell functional $\mathcal{B}$ is obtained, for any level $\ell$, by solving the SDP
\begin{subequations}\label{Eq_Tsirelson_1to2}
\begin{align}
	\max_{\vecP,\,\{u_v\}}&\quad\mathcal{B}\label{Eq_Tsirelson_1to2_obj}\\
	{\rm s.t.}\ \ &\quad\chiDIbc\succeq0\quad\forall\,a,x,\label{Eq_Tsirelson_1to2_psd}\\
	&\quad\sum_a\chiDIbc=\sum_a\chilbb[\rho^{\ttBC}_{a|x'}]\ \ \ \forall\,x\neq x',\label{Eq_Tsirelson_1to2_ns}\\
	&\quad\sum_{a}\chiDIbc_{\tr}=1,\label{Eq_Tsirelson_1to2_norm}
\end{align}
\end{subequations}
where the consistency constraint of \eq{Eq_Tsirelson_1to2_ns} is the lift of the condition $\sum_a\raxbc=\rbc$ [cf.\ the discussion below \eq{Eq_assemblage_1to2}], and the probabilities $\vecP$ enter the matrices through the identification of \eq{Eq_observable_1to2}. Applying \eq{Eq_Tsirelson_1to2} to the Mermin and the Svetlichny inequalities recovers the Tsirelson bounds $4$ and $4\sqrt{2}$ in this scenario as well, already at the first level in both cases. The values are collected in Table~\ref{Table_Tsirelson_1to2}.

\begin{table}[t]
\caption{Upper bounds on the Tsirelson bounds of the Mermin and the Svetlichny inequalities in the 1-steer-2 scenario, computed from the SDP of \eq{Eq_Tsirelson_1to2} at level $\ell=1$. In both cases the first-level bound agrees with the Tsirelson bound to within $10^{-7}$.}
\label{Table_Tsirelson_1to2}
\begin{ruledtabular}
\begin{tabular}{lccc}
	Bell inequality & Local & AMM & Tsirelson\\
	 & bound & bound & bound\\
	\hline
	Mermin, \eq{Eq_Mermin} & $2$ & $4.0000$ & $4$\\
	Svetlichny, \eq{Eq_Svetlichny} & $4$ & $5.6569$ & $4\sqrt{2}$\\
\end{tabular}
\end{ruledtabular}
\end{table}

In contrast with the 2-steer-1 scenario, however, post-quantum steering cannot occur here: since there is a single steering party, every assemblage $\{\raxbc\}_{a,x}$ satisfying positivity and the consistency condition $\sum_a\raxbc=\rbc$ admits a quantum realization of the form of \eq{Eq_assemblage_1to2}, irrespective of the bipartite structure of the steered system~\cite{Gisin89,Hughston93} (see also Ref.~\cite{Sainz17}). Whether the characterization provided by the AMMs converges, as $\ell\to\infty$, to the quantum set therefore remains an open question in this scenario, just as it does in the bipartite one~\cite{CBLC16,CBLC18}.

\subsection{Lower bounds on steering robustness}
\label{Sec_SRDI_1to2}

Applying the completely positive map underlying the AMM construction to the constraints of \eq{Eq_SR_1to2}, and relaxing the resulting conditions to the device-independent matrices, following step by step the derivation of Sec.~\ref{Sec_SRDI_2to1}, we arrive at the SDP
\begin{subequations}\label{Eq_SRDI_1to2}
\begin{align}
	\SRDIonetwo(\Bobs)=\min_{\{u_v\}}&\quad\sum_\lambda\chilbb[\sigma^{\ttBC}_\lambda]_{\tr}-1\label{Eq_SRDI_1to2_obj}\\
	{\rm s.t.}\ \ &\quad\sum_\lambda D(a|x,\lambda)\,\chilbb[\sigma^{\ttBC}_\lambda]\succeq\chiDIbc\label{Eq_SRDI_1to2_lhs}\\
	&\hspace{11em}\forall\,a,x,\nonumber\\
	&\quad\chilbb[\sigma^{\ttBC}_\lambda]\succeq0\quad\forall\,\lambda,\label{Eq_SRDI_1to2_psd}\\
	&\quad\text{Eqs.~(\ref{Eq_Tsirelson_1to2_psd})--(\ref{Eq_Tsirelson_1to2_norm})},\label{Eq_SRDI_1to2_cons}\\
	&\quad\mathcal{B}=\Bobs,\label{Eq_SRDI_1to2_data}
\end{align}
\end{subequations}
where each $\chilbb[\sigma^{\ttBC}_\lambda]$ is built from the same operator set as $\chiDIbc$, and hence contains, in particular, the analogs of the cross moments involving both $B_{b|y}$ and $C_{c|z}$, but with all of its entries treated as unknowns. This reflects the fact that the hidden states $\sigma^{\ttBC}_\lambda$ of \eq{Eq_LHS_1to2} may carry arbitrary correlations between Bob and Charlie. In the binary scenario, the hidden variable $\lambda$ runs over the $2^2=4$ deterministic strategies of Alice. In complete analogy with \eq{Eq_SR_chain_2to1}, for any quantum realization compatible with the observed data, we have
\begin{equation}\label{Eq_SR_chain_1to2}
	\SR(\{\raxbc\})\geq\SRDIonetwo(\Pobs)\geq\SRDIonetwo(\Bobs),
\end{equation}
where $\SRDIonetwo(\Pobs)$ denotes the optimum obtained upon replacing the constraint of \eq{Eq_SRDI_1to2_data} by the full data-matching condition $P(a,b,c|x,y,z)=\Pobsabcxyz$ for all $a,b,c,x,y,z$.

\begin{figure}[t]
\centering
\subfigure[]{\includegraphics[width=0.85\linewidth]{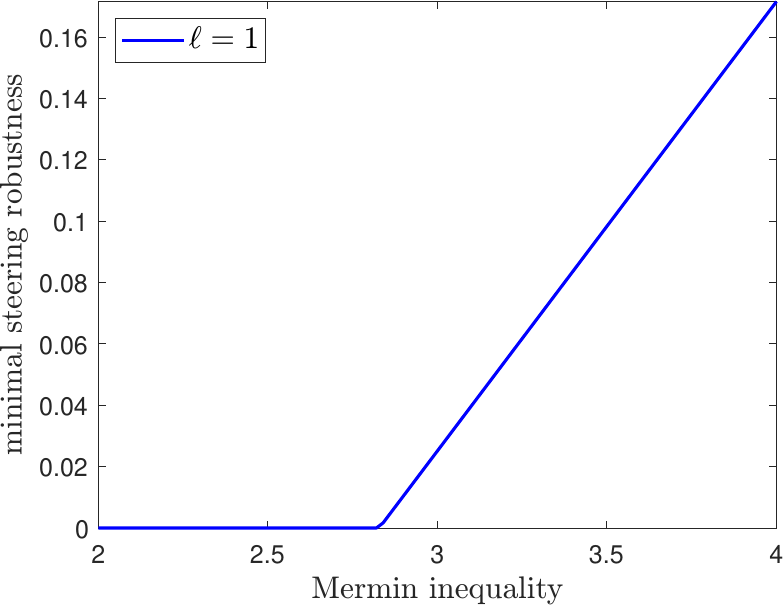}\label{Fig_SR_1to2_Mermin}}\\
\subfigure[]{\includegraphics[width=0.85\linewidth]{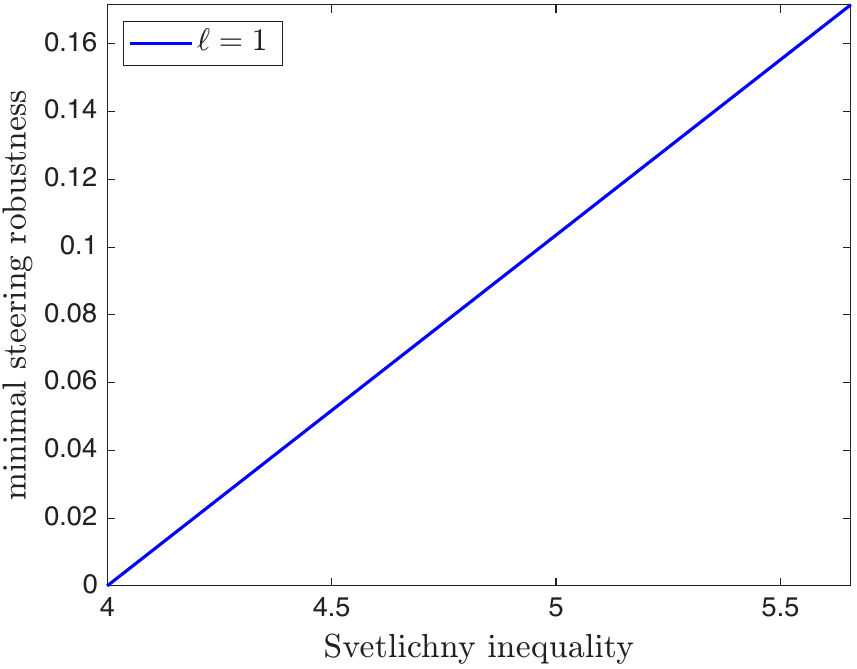}\label{Fig_SR_1to2_Svet}}
\caption{Device-independent lower bounds on the steering robustness in the 1-steer-2 scenario, obtained from the SDP of \eq{Eq_SRDI_1to2}, as functions of the observed violation of (a) the Mermin inequality and (b) the Svetlichny inequality. In panel (a) the bound stays at zero up to $\Bobs=2\sqrt{2}$, for the reason given below \eq{Eq_Mermin_reduced_1to2}.}
\label{Fig_SR_1to2}
\end{figure}

We have solved the SDP of \eq{Eq_SRDI_1to2} for the Mermin and the Svetlichny functionals, varying $\Bobs$ between the corresponding local bound and Tsirelson bound. The results are shown in \fig{Fig_SR_1to2}. In both cases the bound increases linearly with $\Bobs$ over most of the range and reaches $3-2\sqrt{2}\approx0.17$ at the corresponding Tsirelson bound.
For both functionals the second level does not improve on the first, so the first level is already converged.

The two functionals differ in where the bound starts to rise. For the Svetlichny inequality it becomes nonzero immediately above the local bound $\Bobs=4$. For the Mermin inequality, it stays zero until $\Bobs=2\sqrt{2}$, as explained below. If Alice's assemblage admits the local hidden-state model of \eq{Eq_LHS_1to2}, her outcomes are deterministic given $\lambda$, say $A_1=\alpha_1$ and $A_2=\alpha_2$ with $\alpha_1,\alpha_2\in\{\pm1\}$. The Mermin functional of \eq{Eq_Mermin} then reduces to
\begin{equation}\label{Eq_Mermin_reduced_1to2}
	\alpha_1\big(\ew{B_1C_1}-\ew{B_2C_2}\big)-\alpha_2\big(\ew{B_1C_2}+\ew{B_2C_1}\big),
\end{equation}
a CHSH functional for Bob and Charlie, whose quantum maximum is $2\sqrt{2}$. An unsteerable assemblage can therefore produce any Mermin value up to $2\sqrt{2}$, and no bound is possible below that point.

Both bounds are tight at maximal violation. Take the pure GHZ state and let Alice measure $A_1=\sigma_x$ and $A_2=\sigma_y$. With the measurements of Bob and Charlie listed in Table~\ref{Table_endpoints_1to2}, these settings saturate the Mermin and the Svetlichny inequality, respectively. The assemblage, and hence its steering robustness, depends only on the state and on Alice's measurements, and solving \eq{Eq_SR_1to2} for the resulting assemblage gives $\SR(\{\raxbc\})\approx 3-2\sqrt{2}$ (within the numerical precesion up to $10^{-7}$). This value meets the end point of either curve in \fig{Fig_SR_1to2}.

The computed values agree with closed forms. For the Mermin inequality,
\begin{equation}\label{Eq_SRDI_1to2_Mermin}
	\SRDIonetwo(\Bobs)=\frac{2-\sqrt{2}}{4}\,\big(\Bobs-2\sqrt{2}\big)
\end{equation}
on the range $2\sqrt{2}\leq\Bobs\leq4$, and zero below it. For the Svetlichny inequality,
\begin{equation}\label{Eq_SRDI_1to2_Svet}
	\SRDIonetwo(\Bobs)=\frac{\sqrt{2}-1}{4}\,\big(\Bobs-4\big).
\end{equation}
This is the same expression as \eq{Eq_SRNS_DI_Svet}. Both expressions equal $3-2\sqrt{2}$ at the Tsirelson bound. Sampling $\Bobs$ over the whole range at the first level reproduces both expressions to within $10^{-7}$.

\begin{table*}[t]
\caption{Quantum strategies that attain the two end points of each line in \fig{Fig_SR_1to2}. At the lower end point the assemblage admits the model of \eq{Eq_LHS_1to2}, so $\SR=0$. At the upper end point the assemblage is the same for both inequalities, and its steering robustness is $3-2\sqrt{2}$. In the first row Alice holds $\ket{0}$ and Bob and Charlie share $\ket{\Phi^+}=(\ket{00}+\ket{11})/\sqrt{2}$.}
\label{Table_endpoints_1to2}
\begin{ruledtabular}
\begin{tabular}{llccccc}
Inequality & End point & State & $(A_1,A_2)$ & $(B_1,B_2)$ & $(C_1,C_2)$ & $\SR(\{\raxbc\})$\\
\hline
Mermin & $\M=2\sqrt{2}$ & $\ket{0}\otimes\ket{\Phi^+}$ & $(\sigma_z,-\sigma_z)$ & $(\sigma_z,\sigma_x)$ & $\big(\tfrac{\sigma_z+\sigma_x}{\sqrt{2}},\tfrac{\sigma_z-\sigma_x}{\sqrt{2}}\big)$ & $0$\\
Mermin & $\M=4$ & $\ket{\rm GHZ}$ & $(\sigma_x,\sigma_y)$ & $(\sigma_x,\sigma_y)$ & $(\sigma_x,\sigma_y)$ & $3-2\sqrt{2}$\\
Svetlichny & $\mathcal{S}=4$ & $\ket{000}$ & $(\sigma_z,\sigma_z)$ & $(\sigma_z,\sigma_z)$ & $(\sigma_z,-\sigma_z)$ & $0$\\
Svetlichny & $\mathcal{S}=4\sqrt{2}$ & $\ket{\rm GHZ}$ & $(\sigma_x,\sigma_y)$ & $(\sigma_x,\sigma_y)$ & $\big(\tfrac{\sigma_x-\sigma_y}{\sqrt{2}},\tfrac{\sigma_x+\sigma_y}{\sqrt{2}}\big)$ & $3-2\sqrt{2}$\\
\end{tabular}
\end{ruledtabular}
\end{table*}

Both bounds are tight on the whole range. Table~\ref{Table_endpoints_1to2} lists a quantum strategy for each end point. At the lower end points the assemblage admits an LHS model, so $\SR=0$. At the upper end points $\SR\approx 3-2\sqrt{2}$, as computed above. A convex mixture of the two end-point strategies reproduces every point of the line. 

\subsection{Implications for measurement incompatibility}
\label{Sec_IR_1to2}

For a single steering party, the quantitative relation between steerability and measurement incompatibility established in Ref.~\cite{CBLC16} carries over verbatim, as the argument therein is insensitive to the internal structure of the steered system. Indeed, let $t_0=\IR(\MAA)$; then there exist a measurement assemblage $\{N_{a|x}\}_{x,a}$ and a parent POVM $\{G_\gamma\}_\gamma$ such that $(A_{a|x}+t_0\,N_{a|x})/(1+t_0)=\sum_\gamma D(a|x,\gamma)\,G_\gamma$ for all $a,x$. Defining $\tau_{a|x}=\tr_\tA[(N_{a|x}\otimes\id\otimes\id)\,\rabc]$ and $\sigma^{\ttBC}_\gamma=\tr_\tA[(G_\gamma\otimes\id\otimes\id)\,\rabc]$, one immediately finds that the assemblage $\{(\raxbc+t_0\,\tau_{a|x})/(1+t_0)\}_{a,x}$ admits an LHS model of the form of \eq{Eq_LHS_1to2}, whence
\begin{equation}\label{Eq_IR_1to2}
	\IR(\MAA)\geq\SR(\{\raxbc\}).
\end{equation}
Combining this with \eq{Eq_SR_chain_1to2}, the observed value $\Bobs$ alone thus certifies
\begin{equation}\label{Eq_IR_DI_1to2}
	\IR(\MAA)\geq\SRDIonetwo(\Bobs),
\end{equation}
i.e., the incompatibility of Alice's measurements is lower bounded without characterizing any of the devices. Note that, in contrast with \eq{Eq_IR_2to1}, no square root appears here: with a single steering party, the bound takes exactly the same form as in the bipartite scenario~\cite{CBLC16}.

\section{Conclusion}
\label{Sec_conclusion}

We have brought the assemblage-moment-matrix framework---originally devised for the device-independent quantification of bipartite steerability~\cite{CBLC16,CBLC18}---to the tripartite setting, treating on an equal footing the 2-steer-1 and the 1-steer-2 configurations. In both cases, we constructed SDPs that operate on the observed data alone: they reproduce the Tsirelson bounds of the Mermin and the Svetlichny inequalities exactly, already at the first level of the hierarchy, and they convert any observed violation into a certified lower bound on the steering robustness of the underlying assemblage. These bounds grow linearly with the violation. For the Svetlichny inequality they vanish at the local bound in both scenarios. For the Mermin inequality the 1-steer-2 bound vanishes up to $\M=2\sqrt{2}$, and the 2-steer-1 bound with no-signaling responses vanishes identically. Wherever a bound is nonzero it is tight over its whole range, which we showed by exhibiting quantum strategies at both end points and mixing them.
We further showed that the same data certify measurement incompatibility: for a single steering party, the bipartite relation $\IR\geq\SR$ carries over unchanged, whereas for two steering parties, the larger of the two local incompatibility robustnesses is bounded through the square-root relation of \eq{Eq_IR_DI_2to1}.

Several questions emerge naturally from these results. Our LHS models treat all classical correlations between the steering parties as free resources; devising DI quantifiers of \emph{genuine} multipartite steering~\cite{He13} would require finer-grained hierarchies and appears to us an interesting challenge. The two configurations also behave differently with respect to post-quantum steering~\cite{Sainz15,Sainz17}: in the 2-steer-1 scenario, the AMM characterization is a strict superset of the quantum set, and one may ask which additional constraints---if any---could exclude post-quantum assemblages; in the 1-steer-2 scenario, the convergence of the hierarchy to the quantum set remains open, as it does in the bipartite case~\cite{CBLC16,CBLC18}. Extensions to more parties, to other partitions, and to network scenarios are immediate at the level of the construction; identifying the Bell inequalities for which the resulting steering bounds are tight strikes us as worth pursuing. Finally, in the bipartite setting, tight DI bounds on incompatibility have enabled a form of self-testing of measurements~\cite{CBLC16}; whether the tripartite bounds derived here can play an analogous role deserves further investigation.

\acknowledgements S.-L.~C. acknowledges the support of the National Science and Technology Council (NSTC) Taiwan (Grant No. 115-2628-M-005-001-), National Center for Theoretical Sciences Taiwan (Grant No. 115-2124-M-002-014-), and Center for Quantum Frontiers of Research \& Technology (QFort), National Cheng Kung University, Tainan, Taiwan. X.-H.~W. acknowledges the support of the National Science and Technology Council (NSTC) Taiwan (Grant No. 114-2813-C-005-001-M).

\clearpage
\onecolumngrid
\appendix

\section{Proof of Lemma~\ref{Lem_IR_2to1}}
\label{Sec_App_IR_2to1}

Throughout this appendix we abbreviate
\begin{equation}\label{Eq_App_st}
	s=\IR(\MAA),\qquad t=\IR(\MAB),
\end{equation}
so that the claim of \eq{Eq_IR_2to1} reads
\begin{equation}\label{Eq_App_claim}
	\big[1+\max\{s,t\}\big]^2\ \geq\ (1+s)(1+t)\ \geq\ 1+\SRL(\{\rabxy\}).
\end{equation}
We prove the two inequalities separately. Both proofs use only the definitions of $\IR$ and $\SRL$ recalled in Sec.~\ref{Sec_IR_2to1} and in \eq{Eq_SR_2to1}.

\subsection{Proof of the first inequality}
\label{Sec_App_IR_2to1_first}

Write $m=\max\{s,t\}$. By definition $\IR\geq0$, so all the quantities involved are nonnegative. From $m\geq s$ and $m\geq t$ we obtain
\begin{equation}\label{Eq_App_first}
	1+m\geq1+s>0,\qquad 1+m\geq1+t>0.
\end{equation}
Multiplying these two inequalities between positive numbers yields $(1+m)^2\geq(1+s)(1+t)$. 

\subsection{Proof of the second inequality}
\label{Sec_App_IR_2to1_second}

The strategy is as follows. From the incompatibility decompositions of Alice's and of Bob's measurements, we construct an explicit feasible point of the SDP of \eq{Eq_SR_2to1}. Its objective value then upper bounds $\SRL(\{\rabxy\})$.

\emph{Step 1: unfolding the definition of the incompatibility robustness.} The optimization defining $\IR$ is an SDP whose objective is bounded below and whose sublevel sets are compact, so its optimum is attained. There therefore exist a measurement assemblage $\{N^{\tA}_{a|x}\}_{x,a}$, a parent POVM $\{G^{\tA}_\mu\}_\mu$, and a deterministic response $D(a|x,\mu)$ such that
\begin{equation}\label{Eq_App_JM_A}
	\frac{A_{a|x}+s\,N^{\tA}_{a|x}}{1+s}=\sum_\mu D(a|x,\mu)\,G^{\tA}_\mu\quad\forall\,a,x,
\end{equation}
and likewise, for Bob, a measurement assemblage $\{N^{\tB}_{b|y}\}_{y,b}$ and a parent POVM $\{G^{\tB}_\nu\}_\nu$ with
\begin{equation}\label{Eq_App_JM_B}
	\frac{B_{b|y}+t\,N^{\tB}_{b|y}}{1+t}=\sum_\nu D(b|y,\nu)\,G^{\tB}_\nu\quad\forall\,b,y.
\end{equation}
It is convenient to give the numerators a name,
\begin{equation}\label{Eq_App_tilde_AB}
	\hat{A}_{a|x}\equiv A_{a|x}+s\,N^{\tA}_{a|x},\qquad \hat{B}_{b|y}\equiv B_{b|y}+t\,N^{\tB}_{b|y},
\end{equation}
so that Eqs.~(\ref{Eq_App_JM_A}) and (\ref{Eq_App_JM_B}) become $\hat{A}_{a|x}=(1+s)\sum_\mu D(a|x,\mu)G^{\tA}_\mu$ and $\hat{B}_{b|y}=(1+t)\sum_\nu D(b|y,\nu)G^{\tB}_\nu$. Note that $\hat{A}_{a|x}\succeq0$ and $\hat{B}_{b|y}\succeq0$. Each is a sum of positive semidefinite operators with nonnegative weights.

\emph{Step 2: the cross terms.} We now compare the product $\hat{A}_{a|x}\otimes\hat{B}_{b|y}$ with the original product $A_{a|x}\otimes B_{b|y}$. Expanding \eq{Eq_App_tilde_AB} term by term gives, for all $a,b,x,y$,
\begin{equation}\label{Eq_App_cross}
	\hat{A}_{a|x}\otimes\hat{B}_{b|y}-A_{a|x}\otimes B_{b|y}=s\,N^{\tA}_{a|x}\otimes B_{b|y}+t\,A_{a|x}\otimes N^{\tB}_{b|y}+st\,N^{\tA}_{a|x}\otimes N^{\tB}_{b|y}.
\end{equation}
The three terms on the right-hand side are exactly the cross terms generated by the two mixtures. POVM elements are positive semidefinite by definition. Each of these terms is thus a tensor product of two positive semidefinite operators, and is itself positive semidefinite. The prefactors $s$, $t$, and $st$ are nonnegative. The whole right-hand side of \eq{Eq_App_cross} is therefore positive semidefinite, i.e.,
\begin{equation}\label{Eq_App_cross_psd}
	\hat{A}_{a|x}\otimes\hat{B}_{b|y}\succeq A_{a|x}\otimes B_{b|y}\quad\forall\,a,b,x,y.
\end{equation}
No attempt is made here to interpret the cross terms as an assemblage in their own right. All we retain from them is the operator inequality of \eq{Eq_App_cross_psd}. This is all that will be needed, because the LHS constraint of \eq{Eq_SR_2to1} is itself an inequality.

\emph{Step 3: the states left to Charlie.} Define
\begin{equation}\label{Eq_App_tilde_rho}
	\tilde{\rho}^{\ttC}_{ab|xy}=\tr_{\tAB}\big(\rabc\,\hat{A}_{a|x}\otimes\hat{B}_{b|y}\otimes\id\big)\quad\forall\,a,b,x,y.
\end{equation}
The map $M\mapsto\tr_{\tAB}(\rabc\,M\otimes\id)$ is linear, and it sends positive semidefinite operators to positive semidefinite operators, since $\bra{\phi}\tr_{\tAB}(\rabc\,M\otimes\id)\ket{\phi}=\tr(\rabc\,M\otimes\proj{\phi})\geq0$ for every $\ket{\phi}$. Apply it to the positive semidefinite operator $\hat{A}_{a|x}\otimes\hat{B}_{b|y}-A_{a|x}\otimes B_{b|y}$ of \eq{Eq_App_cross}, and recall the definition of $\rabxy$ in \eq{Eq_assemblage_2to1}. We obtain
\begin{equation}\label{Eq_App_dominates}
	\tilde{\rho}^{\ttC}_{ab|xy}-\rabxy=\tr_{\tAB}\big[\rabc\,\big(\hat{A}_{a|x}\otimes\hat{B}_{b|y}-A_{a|x}\otimes B_{b|y}\big)\otimes\id\big]\succeq0.
\end{equation}
That is, the states of \eq{Eq_App_tilde_rho} dominate the members of the original assemblage.

\emph{Step 4: $\{\tilde{\rho}^{\ttC}_{ab|xy}\}$ has a local hidden-state structure.} Substituting the right-hand sides of Eqs.~(\ref{Eq_App_JM_A}) and (\ref{Eq_App_JM_B}) into \eq{Eq_App_tilde_rho} and using linearity once more,
\begin{equation}\label{Eq_App_LHS_form}
	\tilde{\rho}^{\ttC}_{ab|xy}=(1+s)(1+t)\sum_{\mu,\nu}D(a|x,\mu)\,D(b|y,\nu)\,\sigma_{\mu\nu},\qquad \sigma_{\mu\nu}=\tr_{\tAB}\big(\rabc\,G^{\tA}_\mu\otimes G^{\tB}_\nu\otimes\id\big).
\end{equation}
Here each $\sigma_{\mu\nu}$ is positive semidefinite, by the same positivity property of the map. Moreover, since $\sum_\mu G^{\tA}_\mu=\id$ and $\sum_\nu G^{\tB}_\nu=\id$,
\begin{equation}\label{Eq_App_normalization}
	\sum_{\mu,\nu}\sigma_{\mu\nu}=\tr_{\tAB}(\rabc)=\rc,\qquad\text{so that}\qquad\sum_{\mu,\nu}\tr(\sigma_{\mu\nu})=1.
\end{equation}
Consider now the structure of \eq{Eq_App_LHS_form}. There, the dependence on the inputs and outputs of Alice and of Bob factorizes into the two deterministic responses $D(a|x,\mu)$ and $D(b|y,\nu)$. These are precisely of the form appearing in the LHS model of \eq{Eq_LHS_2to1}. The pair $\lambda=(\mu,\nu)$ plays the role of the hidden variable, and any correlation between Alice's and Bob's outcomes is now mediated by $\lambda$ alone.

\emph{Step 5: a feasible point of the SDP, and conclusion.} Collecting the above, set
\begin{equation}\label{Eq_App_feasible}
	\lambda=(\mu,\nu),\qquad \sigma_\lambda=(1+s)(1+t)\,\sigma_{\mu\nu}.
\end{equation}
These operators are positive semidefinite, and combining \eq{Eq_App_LHS_form} with \eq{Eq_App_dominates} gives
\begin{equation}\label{Eq_App_feasibility}
	\sum_\lambda D(a|x,\lambda)\,D(b|y,\lambda)\,\sigma_\lambda=\tilde{\rho}^{\ttC}_{ab|xy}\succeq\rabxy\quad\forall\,a,b,x,y.
\end{equation}
Hence $\{\sigma_\lambda\}_\lambda$ satisfies both constraints of the SDP of \eq{Eq_SR_2to1}, i.e., it is a feasible point. Its objective value follows from \eq{Eq_App_normalization}:
\begin{equation}\label{Eq_App_objective}
	\sum_\lambda\tr(\sigma_\lambda)-1=(1+s)(1+t)\sum_{\mu,\nu}\tr(\sigma_{\mu\nu})-1=(1+s)(1+t)-1.
\end{equation}
Since $\SRL(\{\rabxy\})$ is the \emph{minimum} of that objective over all feasible points, the value attained by our particular choice can only be larger:
\begin{equation}\label{Eq_App_second}
	\SRL(\{\rabxy\})\leq(1+s)(1+t)-1,
\end{equation}
Adding $1$ to both sides gives exactly the second inequality of \eq{Eq_App_claim}. Together with Sec.~\ref{Sec_App_IR_2to1_first}, this completes the proof of Lemma~\ref{Lem_IR_2to1}. \hfill$\blacksquare$

Two remarks are in order. First, the argument never requires the two robustnesses to be raised to a common value. Keeping $s$ and $t$ distinct throughout is what produces the sharper product form $(1+s)(1+t)$. Second, consider the case $s=t=0$, in which both measurement assemblages are jointly measurable. Then \eq{Eq_App_second} gives $\SRL(\{\rabxy\})\leq0$, so the assemblage is unsteerable. This recovers the qualitative statement that incompatible measurements on at least one steering side are necessary for steering.

\section{The no-signaling simulability robustness of a product assemblage}
\label{Sec_App_product}

Definition~\ref{Def_NS_simulable} applies to any joint measurement assemblage. In the 2-steer-1 scenario Alice and Bob act on their own devices, so the assemblage is always of the product form $\MAAB$. In that case the robustness factorizes.

We prove Proposition~\ref{Prop_IRNS_product}, restated here for convenience as
\begin{equation}\label{Eq_IRNS_product}
	\begin{aligned}
	&1+\IRNS(\MAAB)\\
	&\qquad=\big[1+\IR(\MAA)\big]\big[1+\IR(\MAB)\big].
	\end{aligned}
\end{equation}
We first record the two robustnesses as semidefinite programs. For a single party, $\IR(\{M_{m|w}\})$ is the optimum of
\begin{subequations}\label{Eq_App_IR_sdp}
\begin{align}
	\min_{\{G_\lambda\}}&\quad\frac{1}{d}\sum_\lambda\tr(G_\lambda)-1\label{Eq_App_IR_sdp_obj}\\
	{\rm s.t.}&\quad\sum_\lambda D(m|w,\lambda)\,G_\lambda\succeq M_{m|w}\quad\forall\,m,w,\label{Eq_App_IR_sdp_dom}\\
	&\quad G_\lambda\succeq0\quad\forall\,\lambda,\label{Eq_App_IR_sdp_psd}\\
	&\quad\sum_\lambda G_\lambda=\Big[\frac{1}{d}\sum_\lambda\tr(G_\lambda)\Big]\id,\label{Eq_App_IR_sdp_norm}
\end{align}
\end{subequations}
with $d$ the dimension and $\lambda$ running over the deterministic strategies. The substitution $G_\lambda=(1+t)\tilde{G}_\lambda$ turns the parent POVM of Sec.~\ref{Sec_IR_2to1} into the $\{G_\lambda\}$ above, and the noise term becomes the slack in \eq{Eq_App_IR_sdp_dom}. The factor $1/d$ appears because a POVM sums to the identity, whose trace is $d$. The constraint of \eq{Eq_App_IR_sdp_norm} records the same fact, and it cannot be dropped. Summing \eq{Eq_App_IR_sdp_dom} over $m$ only gives $\sum_\lambda G_\lambda\succeq\id$, and the slack in \eq{Eq_App_IR_sdp_dom} is a valid measurement assemblage exactly when $\sum_\lambda G_\lambda$ is a multiple of the identity. Without \eq{Eq_App_IR_sdp_norm} the program computes the steering robustness of the assemblage that $\{M_{m|w}\}$ generates on a maximally entangled state, which can be strictly smaller than $\IR(\{M_{m|w}\})$. The program for $\IRNS(\{M_{ab|xy}\})$ has the same shape, with $\lambda$ running over the extremal no-signaling boxes $\PNS(a,b|x,y,\lambda)$ and with $d=d_\tA d_\tB$. Restricting to extremal boxes loses nothing, by the decomposition used in \eq{Eq_App_NS_feasible}.

We use one elementary fact about the tensor product. If $X\succeq X'\succeq0$ and $Y\succeq Y'\succeq0$, then
\begin{equation}\label{Eq_App_tensor_order}
	X\otimes Y-X'\otimes Y'=(X-X')\otimes Y+X'\otimes(Y-Y')\succeq0,
\end{equation}
both terms being tensor products of positive semidefinite operators.

\subsection{The upper bound}
\label{Sec_App_product_upper}

Write $s=\IR(\MAA)$ and $t=\IR(\MAB)$. Let $\{G^{\tA}_\mu\}$ and $\{G^{\tB}_\nu\}$ be optimal for the two single-party programs, so that
\begin{equation}\label{Eq_App_prod_opt}
	\sum_\mu\tr(G^{\tA}_\mu)=(1+s)\,d_\tA,\qquad\sum_\nu\tr(G^{\tB}_\nu)=(1+t)\,d_\tB.
\end{equation}
Set $G_{(\mu,\nu)}=G^{\tA}_\mu\otimes G^{\tB}_\nu$ and let the associated box be $\PNS(a,b|x,y,(\mu,\nu))=D(a|x,\mu)D(b|y,\nu)$. Such a box is local, hence no-signaling. Using \eq{Eq_App_tensor_order},
\begin{equation}\label{Eq_App_prod_dom}
	\sum_{\mu,\nu}D(a|x,\mu)D(b|y,\nu)\,G^{\tA}_\mu\otimes G^{\tB}_\nu
	=\Big(\sum_\mu D(a|x,\mu)G^{\tA}_\mu\Big)\otimes\Big(\sum_\nu D(b|y,\nu)G^{\tB}_\nu\Big)\succeq A_{a|x}\otimes B_{b|y}.
\end{equation}
Moreover, \eq{Eq_App_IR_sdp_norm} for each party gives $\sum_{\mu,\nu}G^{\tA}_\mu\otimes G^{\tB}_\nu=(1+s)(1+t)\,\id$, so the product satisfies \eq{Eq_App_IR_sdp_norm} for the joint program as well. This is a feasible point of the program for $\IRNS(\MAAB)$. Its objective value follows from \eq{Eq_App_prod_opt},
\begin{equation}\label{Eq_App_prod_val}
	\frac{1}{d_\tA d_\tB}\sum_{\mu,\nu}\tr\big(G^{\tA}_\mu\otimes G^{\tB}_\nu\big)-1=(1+s)(1+t)-1,
\end{equation}
which gives $1+\IRNS(\MAAB)\leq(1+s)(1+t)$.

\subsection{The lower bound}
\label{Sec_App_product_lower}

The dual of \eq{Eq_App_IR_sdp} reads
\begin{subequations}\label{Eq_App_IR_dual}
\begin{align}
	\max_{\{F_{m|w}\},\,W}&\quad\sum_{m,w}\tr\big(F_{m|w}M_{m|w}\big)-1\label{Eq_App_IR_dual_obj}\\
	{\rm s.t.}&\quad\sum_w F_{\lambda_w|w}\preceq W\quad\forall\,\lambda,\label{Eq_App_IR_dual_con}\\
	&\quad F_{m|w}\succeq0\quad\forall\,m,w,\qquad\tr(W)=1.\label{Eq_App_IR_dual_psd}
\end{align}
\end{subequations}
Here $W=\id/d-Z$, where $Z$ is the multiplier of \eq{Eq_App_IR_sdp_norm}, which can be taken traceless. By \eq{Eq_App_IR_dual_con}, $W$ is positive semidefinite, hence a density operator. Dropping \eq{Eq_App_IR_sdp_norm} from the primal amounts to fixing $W=\id/d$. Taking $G_\lambda$ to be the same large multiple of the identity for every $\lambda$ gives a strictly feasible point of \eq{Eq_App_IR_sdp}, so Slater's condition holds and the two optima agree. Let $(\{F^{\tA}_{a|x}\},W^{\tA})$ and $(\{F^{\tB}_{b|y}\},W^{\tB})$ attain the single-party optima. We claim that $F_{ab|xy}=F^{\tA}_{a|x}\otimes F^{\tB}_{b|y}$, together with $W=W^{\tA}\otimes W^{\tB}$, is feasible for the dual of the joint program. The trace condition holds, since $\tr(W^{\tA}\otimes W^{\tB})=1$.

Fix an extremal no-signaling box $V$. Because $V$ is no-signaling, $\sum_bV(a,b|x,y)=V^{\tA}(a|x)$ does not depend on $y$. Whenever $V^{\tA}(a|x)>0$, the ratio $q(b|y)=V(a,b|x,y)/V^{\tA}(a|x)$ is a conditional distribution for Bob alone, for each $y$. Any such distribution is a convex combination of deterministic ones, $q(b|y)=\sum_\nu r(\nu)\,D(b|y,\nu)$, so \eq{Eq_App_IR_dual_con} for Bob gives
\begin{equation}\label{Eq_App_dual_step1}
	\sum_{y,b}V(a,b|x,y)\,F^{\tB}_{b|y}=V^{\tA}(a|x)\sum_\nu r(\nu)\sum_yF^{\tB}_{\nu_y|y}\preceq V^{\tA}(a|x)\,W^{\tB}.
\end{equation}
Tensoring with $F^{\tA}_{a|x}\succeq0$ on the left and summing,
\begin{equation}\label{Eq_App_dual_step2}
	\sum_{a,b,x,y}V(a,b|x,y)\,F^{\tA}_{a|x}\otimes F^{\tB}_{b|y}\preceq\Big(\sum_{a,x}V^{\tA}(a|x)\,F^{\tA}_{a|x}\Big)\otimes W^{\tB}.
\end{equation}
The marginal $\{V^{\tA}(a|x)\}$ is a conditional distribution for Alice alone, so the same decomposition applies once more, now through \eq{Eq_App_IR_dual_con} for Alice. Since $W^{\tB}\succeq0$, the right-hand side of \eq{Eq_App_dual_step2} is therefore at most $W^{\tA}\otimes W^{\tB}$, which is the dual constraint of the joint program. The product certificate is thus feasible, and its objective value is
\begin{equation}\label{Eq_App_dual_val}
	\sum_{a,b,x,y}\tr\big[(F^{\tA}_{a|x}\otimes F^{\tB}_{b|y})(A_{a|x}\otimes B_{b|y})\big]-1=(1+s)(1+t)-1.
\end{equation}
Weak duality gives $1+\IRNS(\MAAB)\geq(1+s)(1+t)$, which together with Sec.~\ref{Sec_App_product_upper} proves \eq{Eq_IRNS_product}. \hfill$\blacksquare$

Two remarks. First, the argument uses no property of $V$ beyond the no-signaling conditions, and it does not use the number of settings, the number of outcomes, or the dimensions. Second, the upper bound of Sec.~\ref{Sec_App_product_upper} is attained already with local boxes, so \eq{Eq_IRNS_product} also equals the robustness obtained when the post-processing is restricted to that smaller class. A consequence is that Corollary~\ref{Cor_IRNS_2to1} and \eq{Eq_IR_SRNS_weak} carry the same bound whenever the steering parties measure locally, which is the only case arising here.

\section{A direct proof of the outer inequality of Corollary~\ref{Cor_IRNS_2to1}}
\label{Sec_App_NS}

The proof uses the map
\begin{equation}\label{Eq_App_Phi}
	\Phi[X]=\tr_{\tAB}\rb{\rabc\,X\otimes\id},
\end{equation}
defined for operators $X$ acting on Alice and Bob's joint space. Three of its properties are needed. It is linear. It maps positive semidefinite operators to positive semidefinite operators, as shown in Step 3 of Appendix~\ref{Sec_App_IR_2to1_second}. Finally, $\Phi[\id]=\rc$ and $\Phi[A_{a|x}\otimes B_{b|y}]=\rabxy$, the latter being the definition of \eq{Eq_assemblage_2to1}.

We also use one fact about the no-signaling polytope, already invoked below \eq{Eq_NS_response}. Every no-signaling box is a convex combination of the extremal boxes $P_{\mbox{\tiny NS}}(a,b|x,y,\lambda)$ that label the hidden variable in the SDP of \eq{Eq_SRNS_2to1}.

Corollary~\ref{Cor_IRNS_2to1} was obtained in Sec.~\ref{Sec_IR_2to1} from Proposition~\ref{Prop_IRNS_product} together with Lemma~\ref{Lem_IR_2to1}. The argument below reaches its outer inequality, $\IRNS(\MAAB)\geq\SRNS(\{\rabxy\})$, directly, without passing through Proposition~\ref{Prop_IRNS_product}. It is the tripartite version of the bipartite argument of Ref.~\cite{CBLC16}, and it leaves that inequality standing independently of the longer proof in Appendix~\ref{Sec_App_product}.

Let $t=\IRNS(\MAAB)$. By Definition~\ref{Def_NS_simulable}, there exist a measurement assemblage $\{N_{ab|xy}\}_{x,y,a,b}$, a parent POVM $\{G_\lambda\}_\lambda$, and no-signaling boxes $V_\lambda(a,b|x,y)$ such that
\begin{equation}\label{Eq_App_NS_mixture}
	\frac{A_{a|x}\otimes B_{b|y}+t\,N_{ab|xy}}{1+t}=\sum_\lambda V_\lambda(a,b|x,y)\,G_\lambda\quad\forall\,a,b,x,y.
\end{equation}
Apply $\Phi$ to both sides and write
\begin{equation}\label{Eq_App_NS_images}
	\tau_{ab|xy}=\Phi[N_{ab|xy}],\qquad \sigma_\lambda=\Phi[G_\lambda].
\end{equation}
Both are positive semidefinite, since $N_{ab|xy}\succeq0$ and $G_\lambda\succeq0$. Using linearity and $\Phi[A_{a|x}\otimes B_{b|y}]=\rabxy$, we obtain
\begin{equation}\label{Eq_App_NS_assemblage}
	\frac{\rabxy+t\,\tau_{ab|xy}}{1+t}=\sum_\lambda V_\lambda(a,b|x,y)\,\sigma_\lambda\quad\forall\,a,b,x,y.
\end{equation}
This already exhibits an admixture of $\{\rabxy\}$ that admits a no-signaling model. It remains to turn it into a feasible point of the SDP of \eq{Eq_SRNS_2to1}.

Rescale by setting $\tilde{\sigma}_\lambda=(1+t)\,\sigma_\lambda$. Since $\tau_{ab|xy}\succeq0$, \eq{Eq_App_NS_assemblage} gives
\begin{equation}\label{Eq_App_NS_dominates}
	\sum_\lambda V_\lambda(a,b|x,y)\,\tilde{\sigma}_\lambda=\rabxy+t\,\tau_{ab|xy}\succeq\rabxy.
\end{equation}
Each $V_\lambda$ is no-signaling, so we may decompose it over the extremal boxes as $V_\lambda(a,b|x,y)=\sum_k q_\lambda(k)\,P_{\mbox{\tiny NS}}(a,b|x,y,k)$ with $q_\lambda(k)\geq0$ and $\sum_k q_\lambda(k)=1$. Substituting into \eq{Eq_App_NS_dominates} and collecting terms gives
\begin{equation}\label{Eq_App_NS_feasible}
	\sum_k P_{\mbox{\tiny NS}}(a,b|x,y,k)\,\hat{\sigma}_k\succeq\rabxy,\qquad \hat{\sigma}_k=\sum_\lambda q_\lambda(k)\,\tilde{\sigma}_\lambda\succeq0.
\end{equation}
The operators $\{\hat{\sigma}_k\}_k$ are therefore feasible for \eq{Eq_SRNS_2to1}. Their objective value is fixed by normalization. Indeed $\sum_k\hat{\sigma}_k=\sum_\lambda\tilde{\sigma}_\lambda=(1+t)\,\Phi[\sum_\lambda G_\lambda]=(1+t)\,\rc$, so that
\begin{equation}\label{Eq_App_NS_objective}
	\sum_k\tr(\hat{\sigma}_k)-1=(1+t)\tr(\rc)-1=t.
\end{equation}
Since $\SRNS(\{\rabxy\})$ is the minimum of that objective over all feasible points, we conclude $\SRNS(\{\rabxy\})\leq t$, which is the outer inequality of \eq{Eq_IRNS_2to1}. \hfill$\blacksquare$

\bibliography{bib_self_testing_states}

\end{document}